%% file: top_ml_paper.tex
\documentclass[a4paper,11pt]{article}
\pdfoutput=1
\usepackage{jheppub}

\usepackage{graphicx}
\usepackage{subfigure}
\usepackage{tabularx, multirow, booktabs}
\newcolumntype{Y}{>{\centering\arraybackslash}X}

\usepackage{hyperref}
\usepackage{top_ml_paper_defs}

\AtBeginDocument{
\heavyrulewidth=.08em
\lightrulewidth=.05em
\cmidrulewidth=.03em
\belowrulesep=.65ex
\belowbottomsep=0pt
\aboverulesep=.4ex
\abovetopsep=0pt
\cmidrulesep=\doublerulesep
\cmidrulekern=.5em
\defaultaddspace=.5em
}

\begin{document}

\title{Jet Constituents for Deep Neural Network Based Top Quark Tagging}

\author[a]{J. Pearkes,}
\author[a,1]{W. Fedorko,\note{Corresponding author.}}
\author[a]{A. Lister}
\author[a]{C. Gay}
\affiliation[a]{Department of Physics and Astronomy, The University of British Columbia, BC, Canada}

\emailAdd{jannicke.pearkes@cern.ch}
\emailAdd{wojtek.fedorko@cern.ch}
\emailAdd{alison.lister@cern.ch}
\emailAdd{colin.gay@cern.ch}

\date{\today}

\abstract{
Recent literature on deep neural networks for tagging of highly energetic jets resulting from top quark decays has focused on image based techniques or multivariate approaches using high-level jet substructure variables. Here, a sequential approach to this task is taken by using an ordered sequence of jet constituents as training inputs. Unlike the majority of previous approaches, this strategy does not result in a loss of information during pixelisation or the calculation of high level features.  The jet classification method achieves a background rejection of 45 at a 50\% efficiency operating point for reconstruction level jets with transverse momentum range of 600 to 2500 GeV and is insensitive to multiple proton-proton interactions at the levels expected throughout Run 2 of the LHC.

}

\maketitle
\flushbottom

\section{Introduction}\label{sec:intro} 
\input{sections/sec-introduction}

\section{Dataset}\label{sec:dataset} 
\input{sections/sec-dataset}

\section{Network Architecture}\label{sec:arch} 
\input{sections/sec-network-architecture}

\newpage
\section{DNN Performance}\label{sec:perf}
\input{sections/sec-performance-analysis}

\section{Conclusions}\label{sec:concl}
\input{sections/sec-conclusions}

\newpage
\clearpage
\bibliography{top_ml_paper}{}

\end{document}

%% file: sections/sec-introduction.tex
The use of boosted top tagging algorithms in both searches and measurements is becoming more prevalent as the LHC datasets are increasing, thus extending the mass reach for searches involving new particles with top quarks in their topology. 
Some recent examples from ATLAS and CMS include the search for new resonances decaying to top pairs in the all-hadronic and semi-leptonic top decay channels~\cite{ttbar_res_ATLAS_13TeV, ttbar_res_ATLAS_8TeV, ttbar_res_CMS_8TeV}, 
searches for vector-like quarks in decays of $T\rightarrow Ht$~\cite{vlq_htx_CMS}, $T\rightarrow Zt$~\cite{vlq_single_zt_CMS} or $B \rightarrow Wt$~\cite{vlq_wbx_ATLAS}, searches for excited $b$ quarks~\cite{excited_b_wt_CMS}, searches for supersymmetric top squarks~\cite{stop_CMS_0l_pub,stop_ATLAS_1l_pub,stop_ATLAS_0l_conf,stop_ATLAS_1l_conf}
and measurements of the differential cross section of top quark pair production in the all-hadronic channel~\cite{ttbar_diff_allhad_ATLAS_13TeV, ttbar_diff_allhad_ATLAS_8TeV}.
These analyses identify boosted top quarks via selection criteria applied to large radius jets (typically with $R$\footnote{\label{coordinates} In this article a cylindrical coordinate system is adopted with the $z$-axis along the beamline. Polar angle is $\theta$, azimuthal angle is $\phi$. Transverse momentum \pt\ is the component of particle momentum in the $x-y$ plane. Pseudorapidity $\eta$ is defined as $\eta=-\ln\left(\tan\left(\frac{\theta}{2}\right)\right)$. Jet clustering parameter $R$ is defined as $R=\sqrt{\Delta y^2 + \Delta\phi^2}$ where $y$ is rapidity.} parameter of 0.8 to 1.2). In addition to the requirements that these jets have high transverse momentum (typically above a couple of hundred GeV), the majority of analyses impose an additional requirement based on the invariant mass of the jet and on explicit requirements on jet substructure variables, such as $\tau_{32}$~\cite{nsubjettiness}, $\sqrt{d_{23}}$ ~\cite{split23} or $D_{2}$~\cite{energycorrelation}. This is done to reject the non-top Quantum Chromodynamics (QCD) background, in which the jets originate from lighter quarks or gluons.
Similar techniques have been applied to the identification of boosted vector bosons, which have also been used in a number of recent results, e.g. Ref.~\cite{vlq_wbx_ATLAS}.

The relative performance of a number of substructure based techniques have been studied in both ATLAS~\cite{atlas_tagging_perf} and CMS~\cite{cms_tagging_perf}. Typical performances for the best top taggers have background rejection ratios of approximately 15 and 5 for signal efficiencies of 50\% and 80\% respectively, for jets around 1~TeV in transverse momentum.

In the past couple of years there have been a number of studies using machine learning techniques to improve the performance of boosted top, $W$, $Z$ or Higgs tagging. 
One of the first papers looked at the use of jet images and techniques derived from computer vision (such as facial recognition algorithms) to gain further insight into the structure of boosted hadronic $W$ boson decays~\cite{jet_image_1}. This study was carried out using Monte Carlo particle-level information and obtained a 20\% improvement in the background rejection, using Fisher's Linear Discriminant~\cite{FischersLD}, over a more traditional N-subjettiness ratio ($\tau_2/\tau_1$).

This idea was further developed in Ref.~\cite{jet_image_11} and applied to the hadronic decay of top quarks where, using a two hidden-layer neural network, a 4\% mistag rate (background rejection factor of 25) was achieved for 60\% signal efficiency in jets with transverse momentum in the range 1100--1200~GeV and jet masses between 130 and 210~GeV.

A later paper extends this jet-image processing, in $W$ boson decays, by including modern deep neural networks (DNN) with convolutional layers~\cite{jet_image_2}. This study was carried out using Monte Carlo particle-level information and obtained a rejection factor of $\sim$30 at 50\% signal efficiency for jets with transverse momenta in the range 250--300 GeV, and a jet mass between 65 and 95 GeV when combining a DNN with high-level features.

Using a mixture of locally-connected and fully-connected nodes in a DNN architecture, applied to $W$ boson identification, another study was able to get a rejection rate of about 70 for a 50\% signal efficiency~\cite{irvine}. This was obtained for jets with transverse momenta between 300 and 400~GeV without any restrictions on the jet mass. This showed similar performance to a more traditional boosted decision tree (BDT) trained on six high-level variables. This study included the effects of pileup and used \textsc{delphes}~\cite{delphes} for detector simulation.

In order to better understand how applicable such techniques are to a real LHC analysis, including systematic uncertainties, the performance of a deep learning algorithm was studied on $W$ boson tagging, using a \textsc{delphes} detector simulation, under variations of the parton shower model~\cite{melbourne}; the performance was found to vary by up to 50\%. 
The use of Generative Adversarial Networks (GAN) on jet images for fast simulation has been explored in~Ref.~\cite{berkley, GAN2}.

Another study on the identification of hadronic decays of boosted top quarks using convolutional neural networks includes the expected impact of a finite detector resolution using \textsc{delphes} assuming similar calorimeter granularity to that of the CMS detector~\cite{deep_top}. A relatively low jet transverse momentum range from 350 to 400~GeV is explored, though a large jet parameter is chosen, $R$=1.5, presumably to ensure the top decay is fully contained. The performance of their best tagger, DeepTop, gives a rejection factor of $\sim$40 for a signal efficiency of 50\%; it slightly out-performs their BDT  based tagger, trained on high-level substructure variables.

A recent paper~\cite{kyle} has taken an alternative look at the problem by using recursive neural networks built upon an analogy between QCD and natural language processing. Variable length sets of four-momenta are used as input to the training. The performance of this method on the identification of boosted $W$ bosons, using calorimeter tower emulation, results in a background rejection factor of $\sim$25 for a 50\% signal efficiency. The ideal rejection factor, using particles as input to the training, is found to be $\sim$70 for the same signal efficiency.  This study looked at jets in the same \pt\ and mass ranges as Ref.~\cite{melbourne} with jet \pt\ between 200 and 500 GeV and jet mass in the range 50--110 GeV.

The study of boosted event shapes for the identification of top, Higgs and vector bosons was studied in~\cite{davis}. In particular a method is described to account for the momentum spread within a sample by boosting into the centre-of-mass frame of the original particle.

Recently another paper looks at the performance of  boosted decision trees and deep neural networks trained on high-level variables for both top and W-tagging. This work achieves a background rejection of $\sim$120 at 50\% signal efficiency in the \pt\ range of 500--1000~GeV and a background rejection of $\sim$30 at 50\% signal efficiency in the \pt\ range of 1000--1500~GeV ~\cite{atlas_bdt_dnn}. 

While much of the literature has so far focused on describing methods, most have used the performance of $W$ boson identification as a benchmark, while only Refs.~\cite{jet_image_11, deep_top, davis, atlas_bdt_dnn} have explored the performance of the algorithms top tagging. 
It is often difficult to compare the relative performance of many of these algorithms due to differing momentum ranges, pre-selection cuts, use of detector simulation and inclusion of pileup.
In addition to top and W-tagging, jet images have been investigated for use in quark/gluon tagging~\cite{quark_gluon_dnn,atlas_qg_tagging}.  There has also been some work to extend these ideas to jet flavour, in particular $b$-jet identification~\cite{irvine_uw, atlas_btag_rnn}.  

\begin{figure}[h!]
\centering
\includegraphics[width=0.32\textwidth]{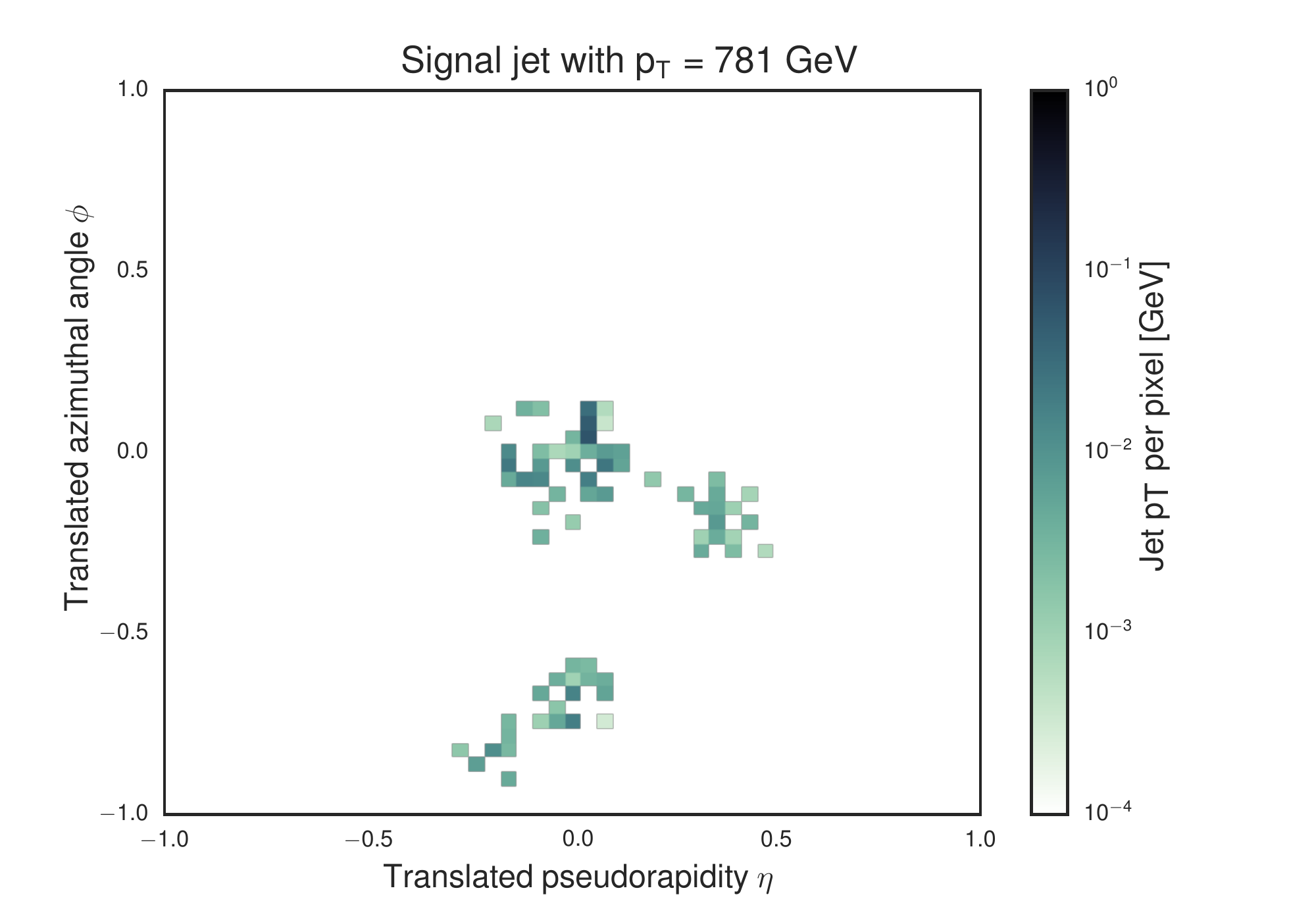}
\includegraphics[width=0.32\textwidth]{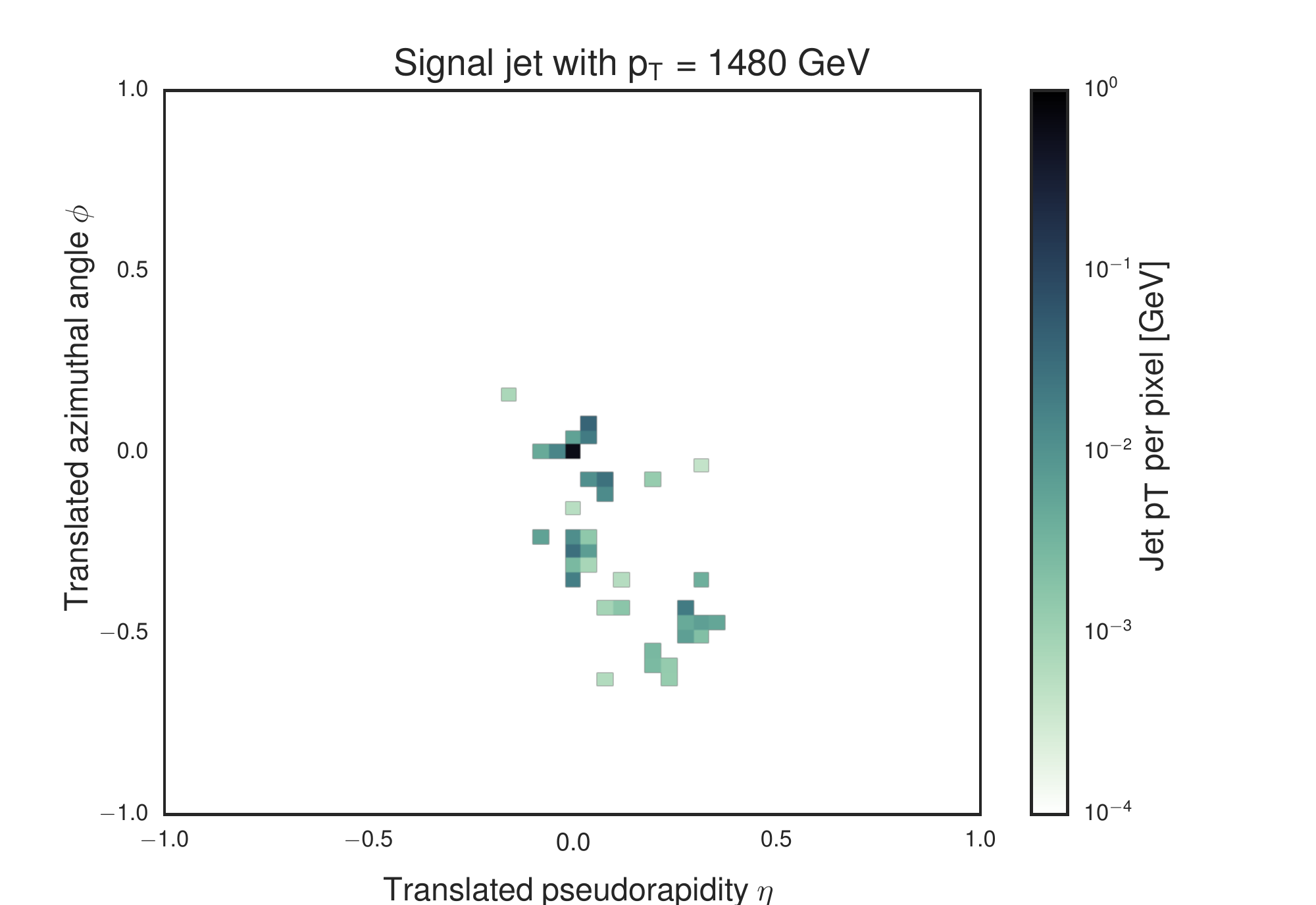}
\includegraphics[width=0.32\textwidth]{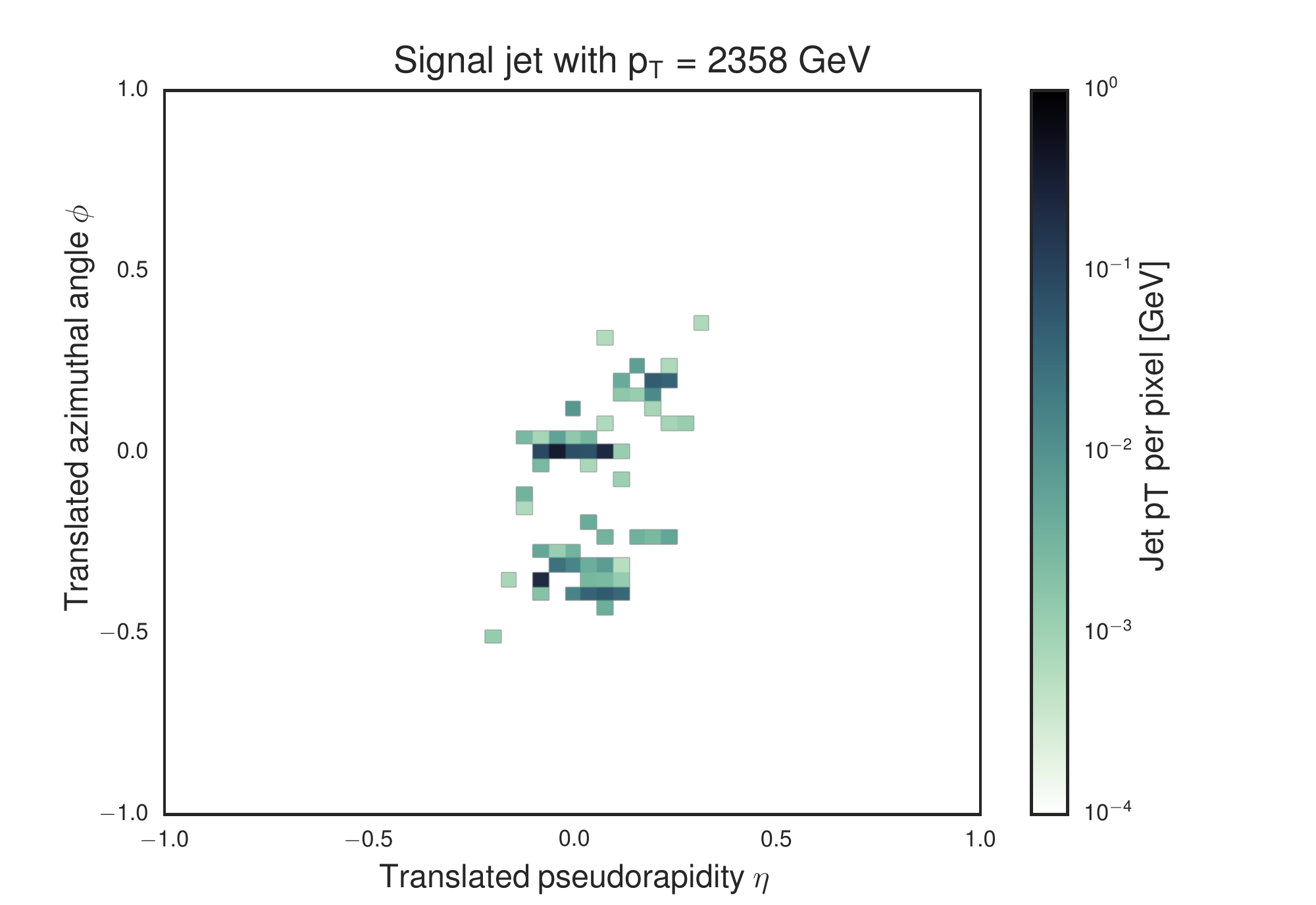}
\includegraphics[width=0.32\textwidth]{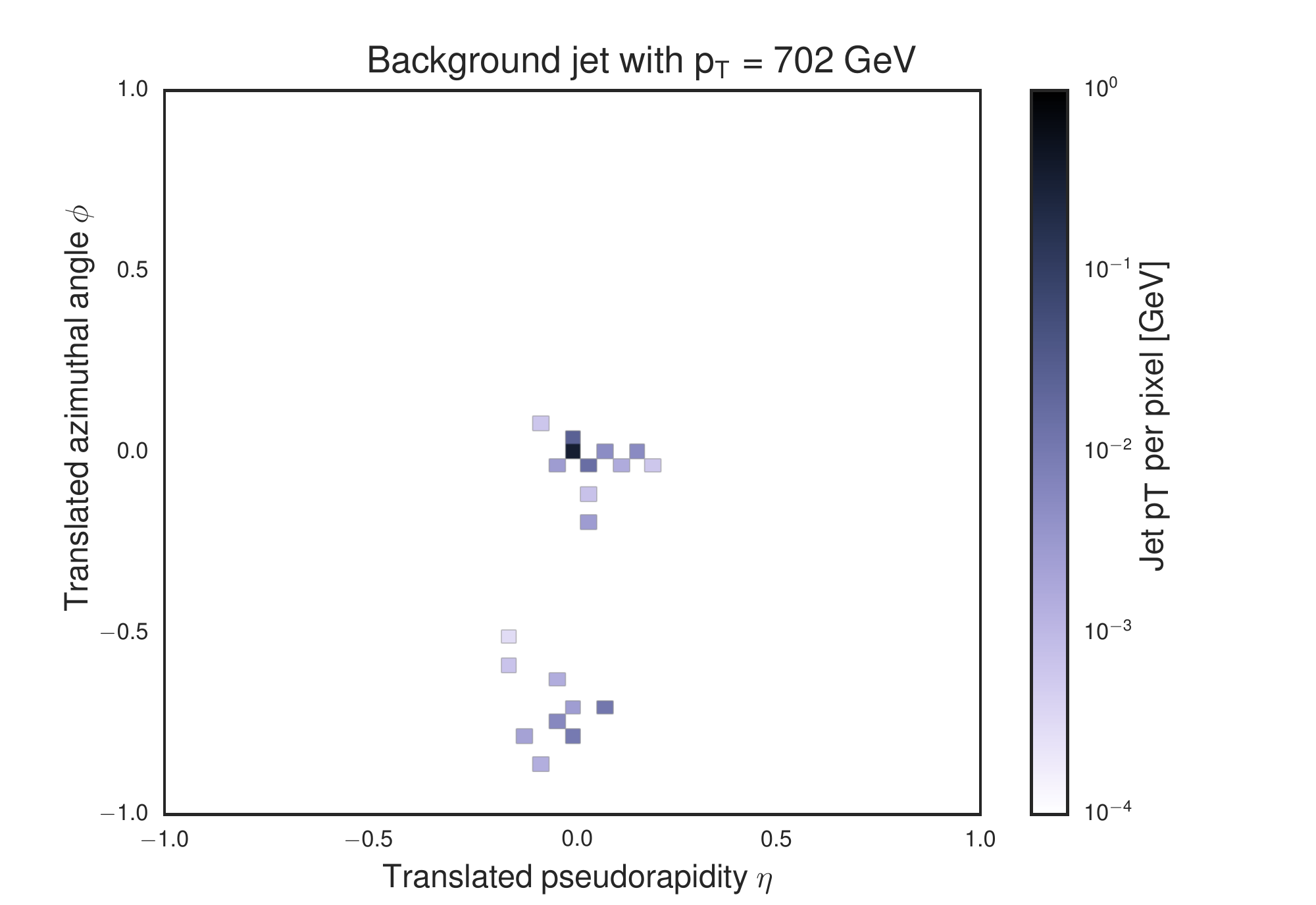}
\includegraphics[width=0.32\textwidth]{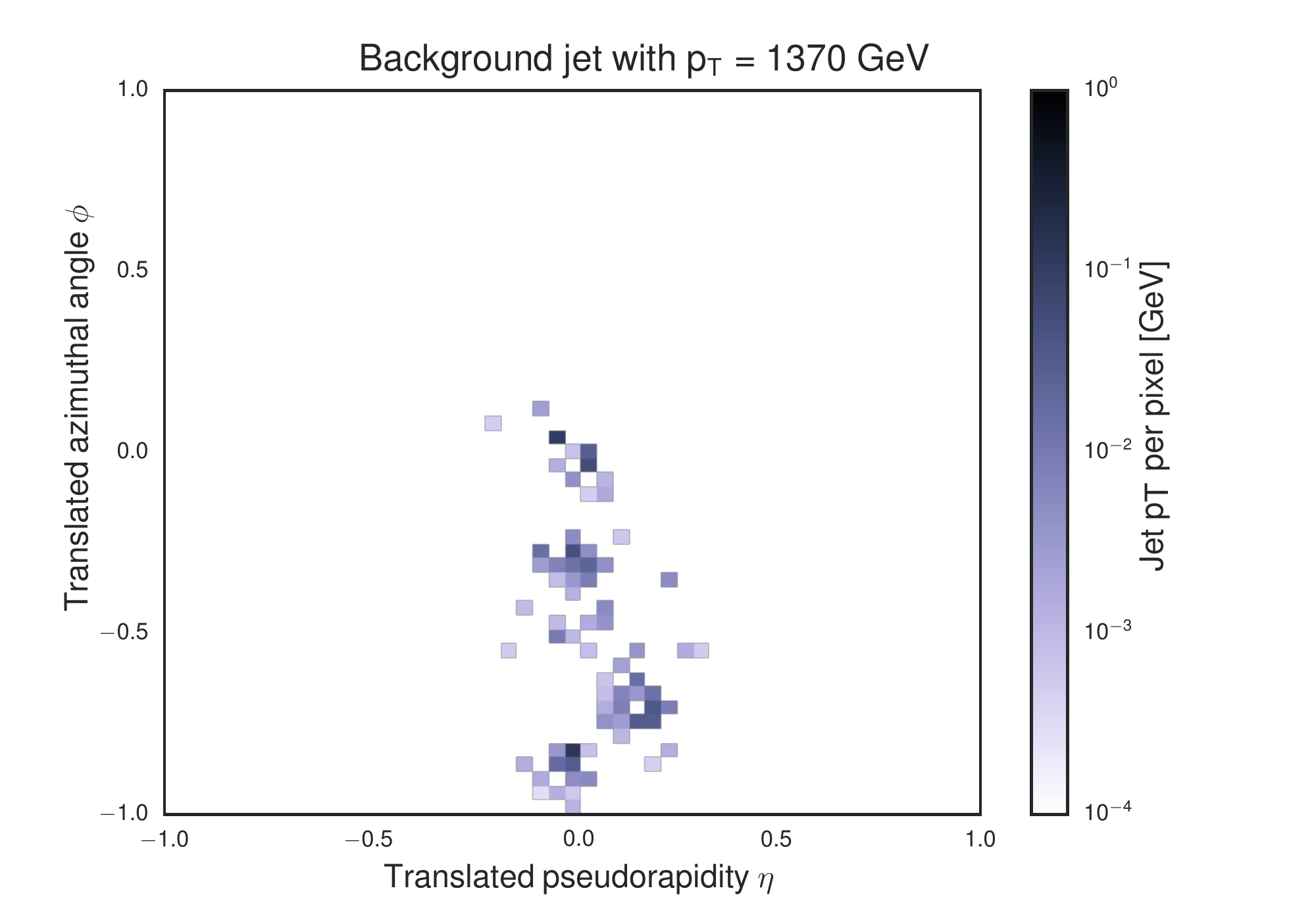}
\includegraphics[width=0.32\textwidth]{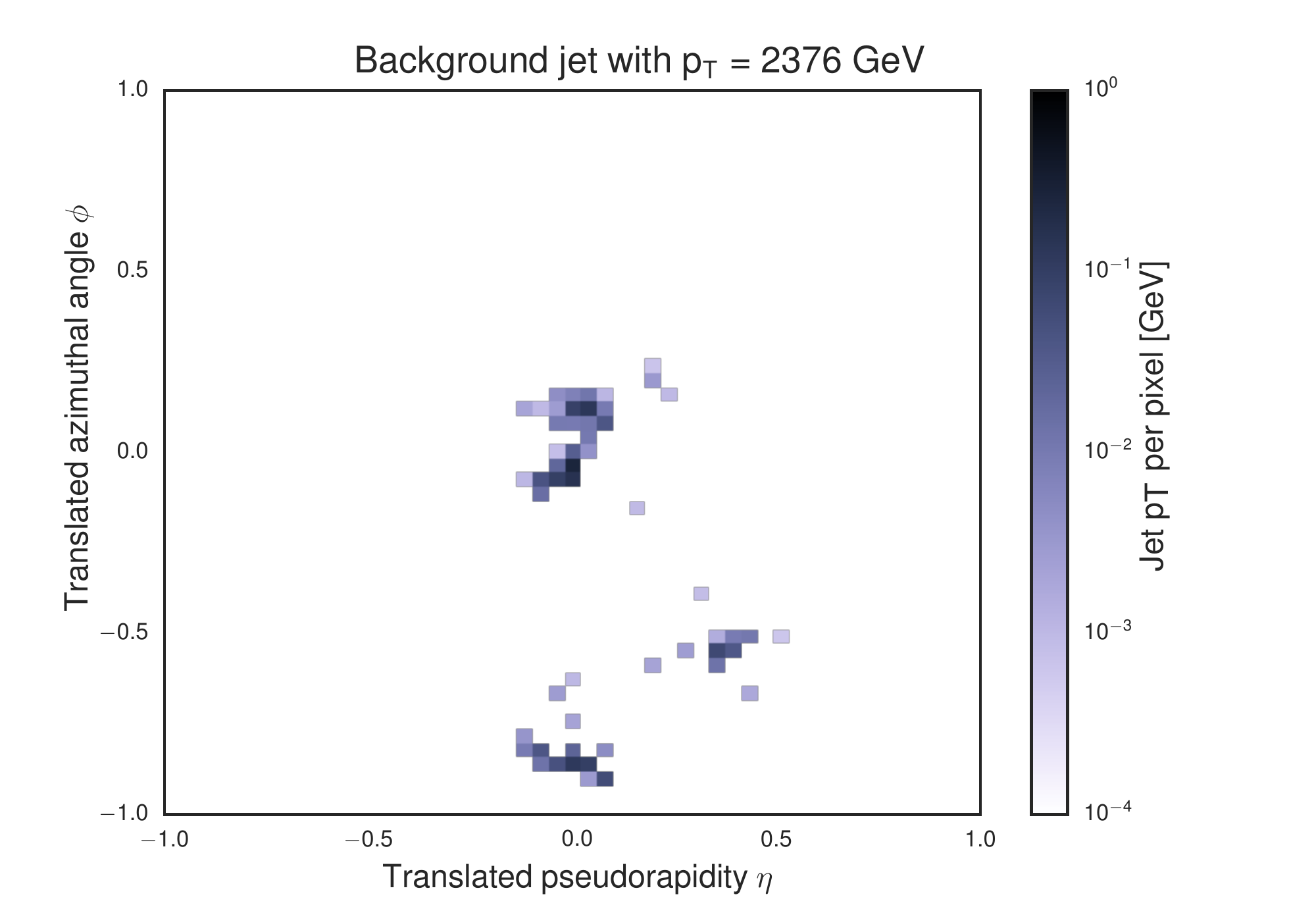}
\caption{Histograms of the fraction of jet \pt\ carried by constituents in $\eta - \phi$ space for examples of signal and background jets. Jets were preprocessed as described in section~\ref{sec:arch}. \label{fig:example_figs}}
\end{figure}

In this paper a method based on a DNN is presented for discriminating top-quark originated jets, henceforth referred to as signal, from jets originating from all other quark flavours and gluons, henceforth referred to as background. The focus of this article are jets with \pt\ above 600 GeV and up to 2500 GeV. In this regime, the $R$=1.0 jets considered are expected to fully contain top quark decays most of the time. 
In Refs.~\cite{jet_image_11,deep_top} deep convolutional neural network-based top tagging methods were developed drawing on the success of these techniques in image recognition. 
However, there are a number of reasons why convolutional networks might not be the optimal architecture for top tagging.
As illustrated in Fig.~\ref{fig:example_figs} in which examples of energy deposits in signal and background jets are shown, the detector activation within a jet is sparse, with most of the detector area within the jet not being activated. 
In addition, no discernible features such as edges, corners or arcs are present. 
Finally, as shown in Ref.~\cite{jet_image_2}, large filter sizes are required in order to achieve competitive performance using convolutional neural networks. 
 
In this article, the performance of fully connected deep neural networks utilising four-vectors of jet constituents is examined. 
The datasets and selection used are described in section~\ref{sec:dataset}, followed by a discussion of the network architecture in section~\ref{sec:arch}.
The data preprocessing techniques applied are described in section~\ref{sec:arch}. 
The performance of the network, the effects of preprocessing, the dependence of the performance on pileup as well as the performance comparison to high-level features is shown in section~\ref{sec:perf}.

%% file: sections/sec-dataset.tex
\subsection{Signal and background modelling}
\label{sec:MC}

The modelling of jets resulting from hadronic top quark decays as well as gluon and light-flavour jets relies on Monte Carlo (MC) simulation. All processes are generated at leading order (LO) at centre-of-mass of 13~TeV using the A14 tune~\cite{A14tune} of the \textsc{pythia} v8.219~\cite{pythia8} event generator with the \textsc{nnpdf23\_lo\_as\_0130\_qed} PDF set, implemented in the \textsc{lhapdf} package~\cite{LHAPDF}. 
The hard scattering process, hadronisation and showering are simulated in a single step.

Samples of Sequential Standard Model \zp\ boson~\cite{zppaper} production at the LHC with pole masses ranging in 32 steps from 1400 to 6360 GeV are generated. 
A centre-of-mass energy cut is applied at 90\% of the \zp\ boson pole mass as well as a cut on the top quark \pt\ proportional to the \zp\ pole mass. 
These cuts are applied to ensure that the top-jet pseudo-rapidity distribution approximately matches that of the background jets.
Only the \zp\ decay to \ttbar\ final state is permitted with each top quark decaying hadronically. 
Similarly, hard QCD ``dijet'' 2 $\rightarrow$ 2 processes, incorporating gluon-gluon, quark-quark and quark-gluon scattering are generated in 32 bins of the outgoing parton transverse momentum ranging from 470 to 2790 GeV. Outgoing partons can be gluons as well as all quark flavours except for top. Only light flavour quarks are treated as massless in the matrix element. A large sample of inelastic, non-diffractive soft QCD events, commonly referred to as minimum bias is also generated.

The detector response is simulated using the \textsc{delphes} v3.4.0 suite~\cite{delphes} using the default emulation of the CMS detector. 
Minimum bias events are overlaid on the hard scattering process to mimic pileup in which multiple $pp$ collisions occur within a single LHC bunch crossing. 
Three scenarios are simulated. 
In the first, no additional $pp$ interactions are added. 
In the other two, a random number of additional $pp$ collisions is overlaid; 
in one case the number is drawn from a Poisson distribution with mean of 23, approximately mimicking the pileup conditions of the LHC during the 2016 data-taking; 
the other case uses a Poisson distribution with a mean of 50, anticipating the pileup conditions expected at the end of the LHC Run 2. The case where an average of 23 pileup interactions are overlaid is referred to as the LHC 2016 pileup scenario.

\subsection{Jet Selection}
\label{jet_selection} 
Large radius jets are formed from  \textsc{delphes} energy-flow objects that emulate the CMS particle-flow algorithm~\cite{cms_pflow_1,cms_pflow_2}.
The anti-$\mathrm{k_T}$ algorithm~\cite{antikt} implemented by the \textsc{FastJet} package~\cite{fastjet} with radius parameter $R$=1.0 is employed. A trimming procedure~\cite{trimming} is applied, where jet constituents are re-clustered into ``subjets'' using the $\mathrm{k_T}$ algorithm~\cite{kt} with radius parameter $R$=0.2 and constituents that belong to subjets carrying less then 5\% of the jet transverse momentum are removed. Jet four-vectors are calculated using the remaining constituents. No further calibration or pileup subtraction steps are applied. For the simulation run with no pileup added, the jet finding and trimming is also performed on all stable particles output by the generator to evaluate the performance without detector effects.

Signal jets were truth matched such that the $\Delta R=\sqrt{\Delta \eta^2 + \Delta\phi^2}$ between a hadronically decaying top quark and the large radius jet was less than~0.75. In addition, jets were selected to have ${\eta \leq 2.0}$ and a jet \pt\ between 600 and 2500 GeV. After this pre-selection, the generated jets were subsampled in \pt\ and $\eta$ to achieve a flat distribution in \pt, and a signal matched distribution in $\eta$. 
This step was taken to prevent the deep neural network from learning the underlying \pt\ and $\eta$ distributions of the generated signal and background jets.  This selection resulted in approximately 7.5 million jets (3.75 million signal jets and 3.75 million background jets). 
An additional independent set of signal and background samples was also simulated and used for evaluating the performance at high background rejections. After the same selection, this test set contains 11 million jets, evenly split between signal and background.

\subsection{Training, Validation and Test Samples}
\label{samples_tvt}
The 7 million jet sample was divided into training, validation and test sets in an 80\%, 10\%, 10\% split. 
Decisions about which network architecture and preprocessing techniques were to be used were made by evaluating the best performance on the validation set. 
The test subset was used for the initial performance analysis. 
The evaluation of the network performance at operating points corresponding to high background rejection justified the creation of the additional test set of 11 million jets. The sample is divided in 6 batches of 1.9 million jets each. In the network performance results quoted in section~\ref{sec:perf} the first four batches (comprising 7.5 million jets) of this large test sample were used. To ascertain the impact of the size of the test set on the quoted results, the performance metrics of the best performing network were evaluated on 15, 4-batch subsamples of the test set. This evaluation was performed only for the best performing network in the LHC 2016 pileup scenario due to computational constraints.

%% file: sections/sec-network-architecture.tex
\begin{figure}[h!]
\centering
\includegraphics[width=0.8\textwidth]{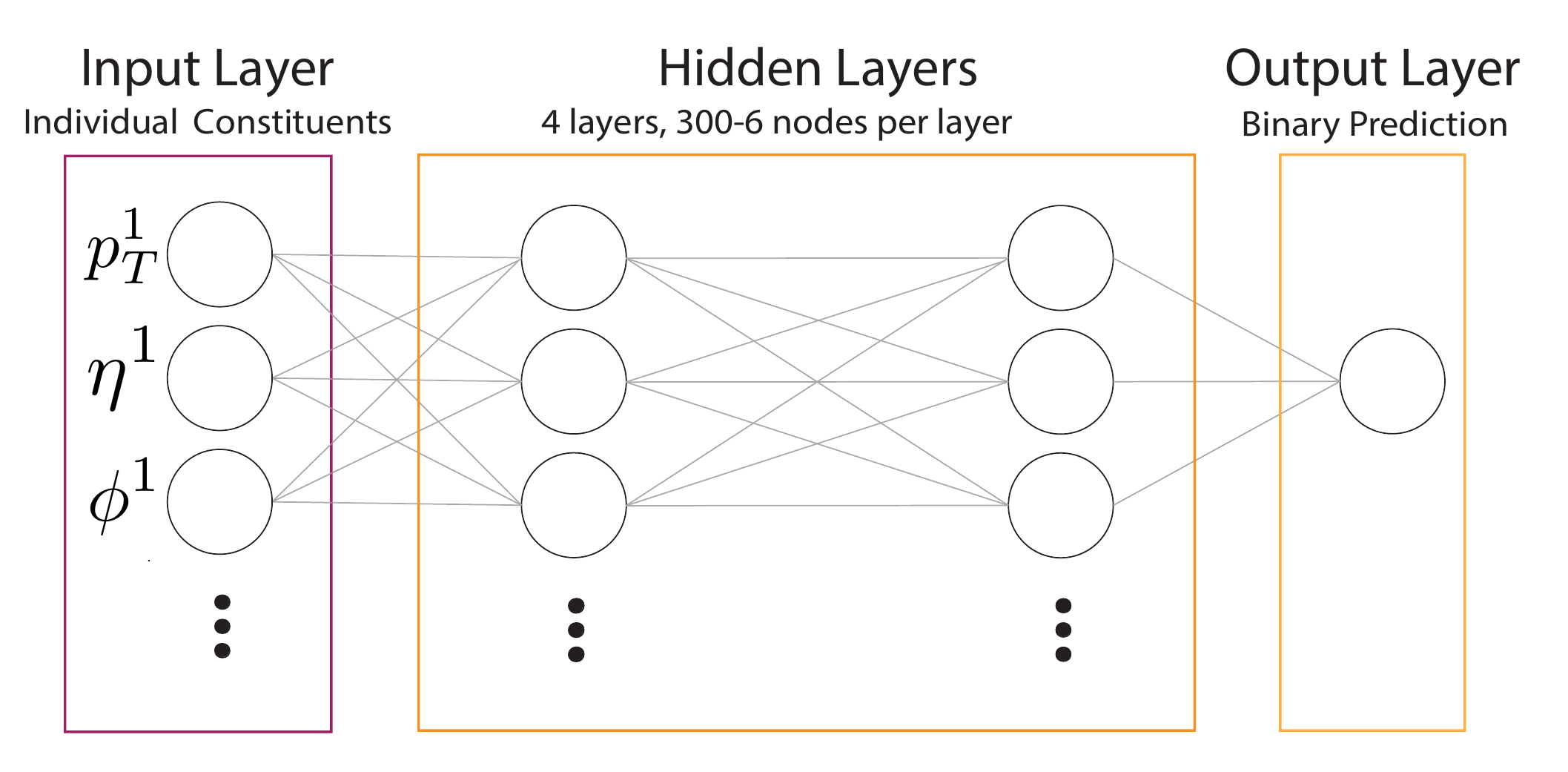}
\caption{Schematic of overall network architecture used.}\label{fig:architecture}
\end{figure}

The networks studied here were implemented using the \textsc{Keras} suite~\cite{keras} with the \textsc{Theano} \cite{theano} backend. 
The input layer of the network consists of a vector of jet constituent \pt, $\eta$ and $\phi$ coordinates. 
The network depth and number of nodes per layer were tuned manually, exploring a space between 4-6 layers and 40-1000 nodes per layer. \textsc{ReLu} activation~\cite{relu} was used for the hidden layers while a sigmoid is used for the output node. 
The network was trained with the \textsc{Adam} optimiser~\cite{adam} for a maximum of 40 epochs. 
Early stopping with a patience parameter of 5 epochs on the loss in the validation set was used. 
The model used for evaluating the performance on the test set is the model with the best performance (lowest binary cross-entropy loss) on the validation set. 
This method prevents overtraining by freezing the model once performance on the validation set begins to decrease. 
The final chosen network architecture consists of 4 hidden layers, with 300, 102, 12 and 6 nodes per layer.    
Figure~\ref{fig:architecture} shows a schematic of the overall network architecture used in this study.

\subsection{Preprocessing}\label{sec:preprocessing}

The key idea behind preprocessing the jets is that, by incorporating domain specific knowledge about the jet physics, the dimensionality of the problem can be reduced. The preprocessing steps were inspired by previous papers~\cite{jet_image_2,irvine,melbourne,deep_top} and determined through a series of studies. Jets are scaled, translated, rotated and flipped.

First, the \pt\ of all jet constituents is scaled by $1/1700$ to ensure that the majority of jet constituents have a \pt\ approximately between zero and one. This ensures that the value of the input nodes corresponding to the \pt\ of the jet constituents are roughly within the same order of magnitude as the input nodes corresponding to the $\eta$ and $\phi$ of the constituents. 

Next, jets are translated in $\eta$ and $\phi$ according to equations~\ref{etatrans} and~\ref{phitrans} so that their primary - highest \pt\ - subjet is centred about $(0,0)$, that is the $y$ and $z$ components of the primary subjet are 0. 
In these and subsequent equations, subjet subscript 0 indicates the primary subjet, subscript 1(n) indicates the subjet with second(n$^{th}$) highest \pt.
\begin{equation}
\eta'_{\mathrm{constituent\,} n} = \eta_{\mathrm{constituent\,} n} - \eta_{\mathrm{subjet\,} 0}
\label{etatrans}
\end{equation}
\begin{equation}
\phi'_{\mathrm{constituent\,} n} = \phi_{\mathrm{constituent\,} n} - \phi_{\mathrm{subjet\,} 0}
\label{phitrans}
\end{equation}

Then, unlike in previous studies~\cite{jet_image_2,irvine,melbourne,deep_top}, the rotations are not performed directly in the $\eta$-$\phi$ plane as it results in a loss of jet mass information, as shown in Refs.~\cite{jet_image_2, deep_top}. 
Instead, a rotation angle $\vartheta$ about the x-axis is computed as shown in equation~\ref{rotangle}. The $y$ and $z$ components of all constituents are then transformed as shown in equations~\ref{pytrans} and~\ref{pztrans}. 
This transformation results in the second highest \pt\ subjet being directly along the negative y-axis.
As this is a Lorentz transformation all the invariants such as the jet mass are preserved.

\begin{equation}
\vartheta = \tan^{-1}(\frac{p_{y,\, \mathrm{subjet\,} 1}}{p_{z,\, \mathrm{subjet\,} 1}})+\frac{\pi}{2}
\label{rotangle}
\end{equation}
\begin{equation}
p'_{y,\, \mathrm{constituent\,} n} = p_{y,\, \mathrm{constituent\,} n} \cos \vartheta - p_{z,\, \mathrm{constituent\,} n} \sin \vartheta
\
\label{pytrans}
\end{equation}
\begin{equation}
p'_{z,\, \mathrm{constituent\,} n} = p_{y,\, \mathrm{constituent\,} n} \sin \vartheta - p_{z,\, \mathrm{constituent\,} n} \cos \vartheta
\label{pztrans}
\end{equation}

\begin{figure}[ht!]
\centering
\includegraphics[width=0.48\textwidth]{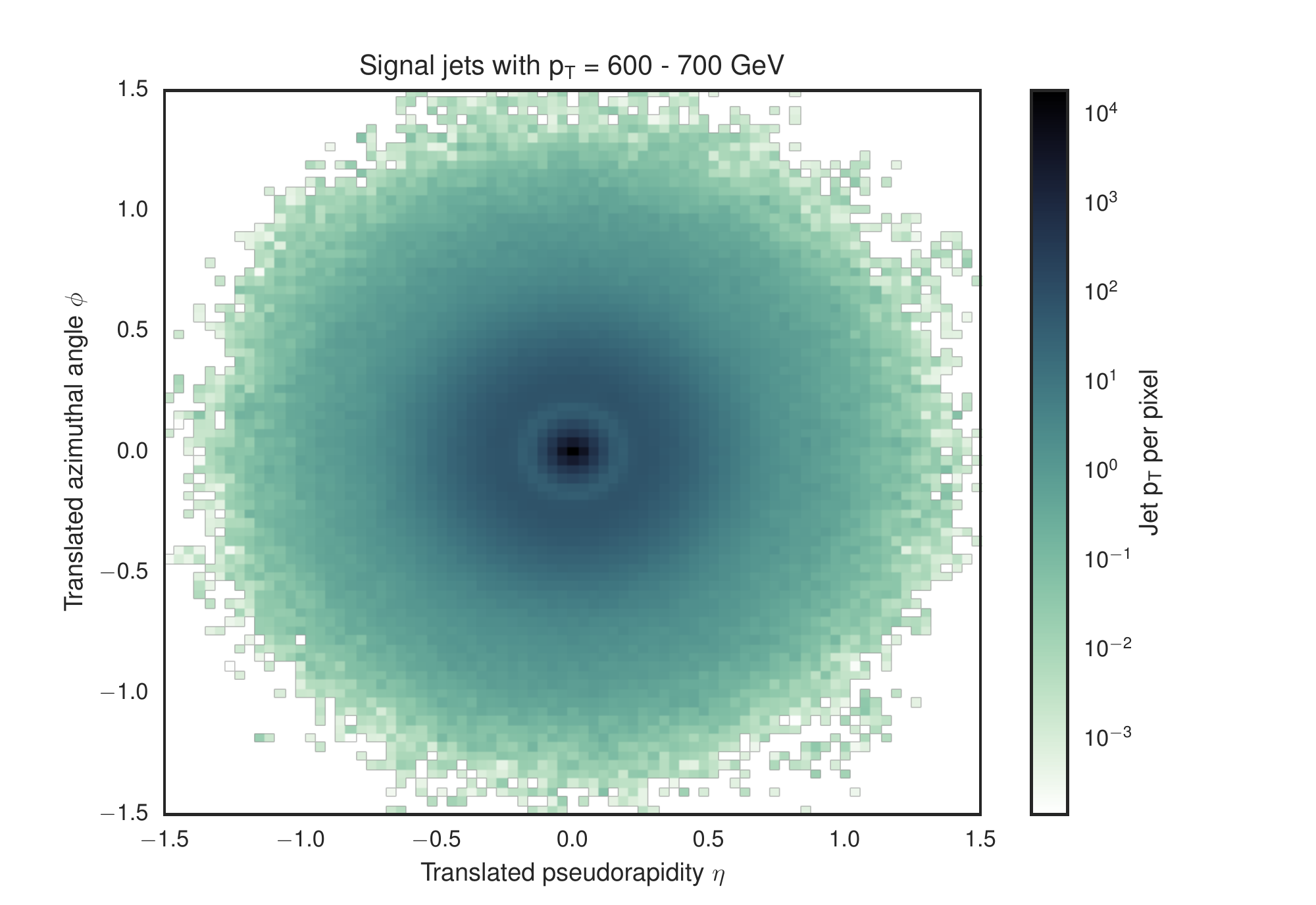}
\includegraphics[width=0.48\textwidth]{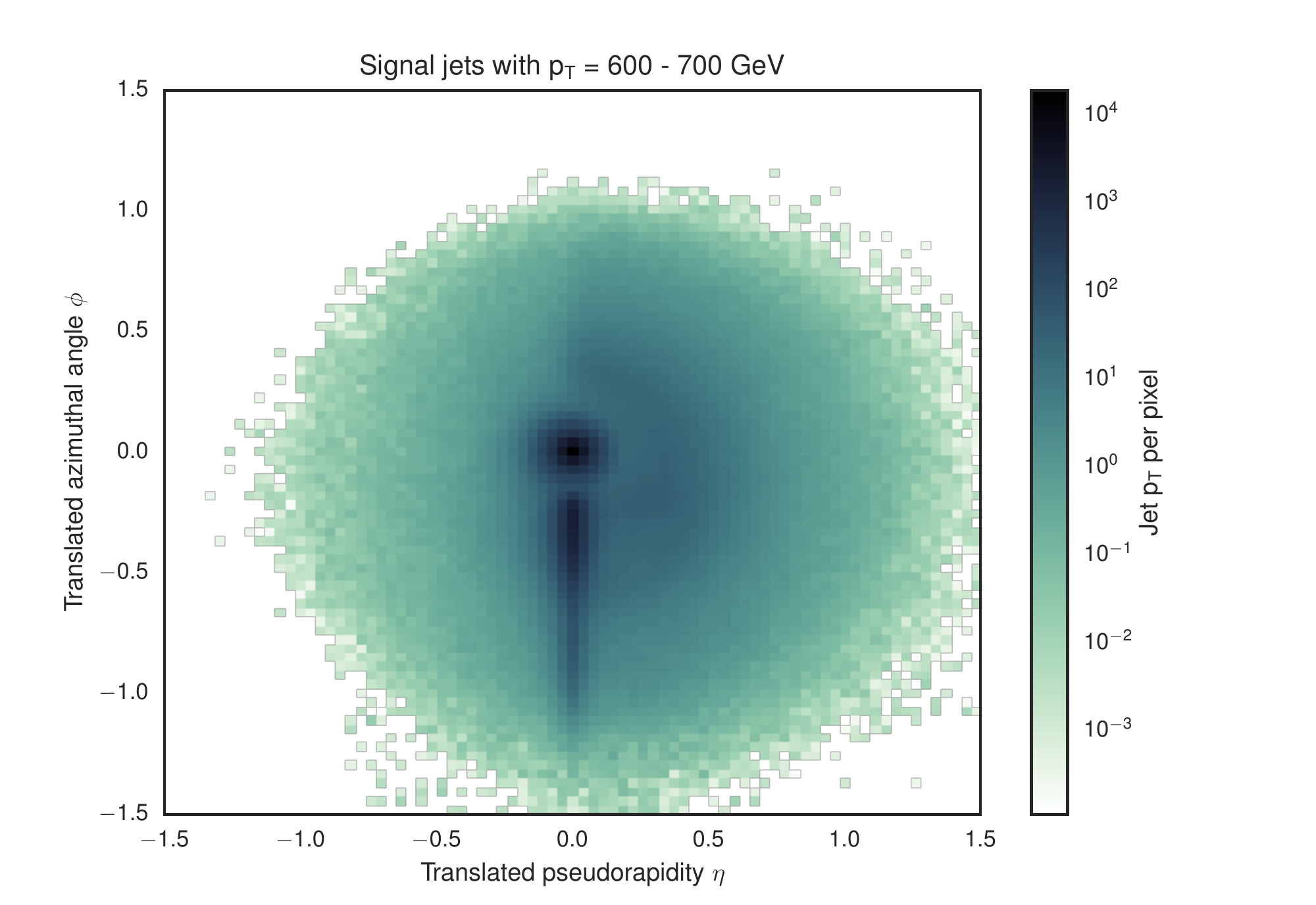}
\includegraphics[width=0.48\textwidth]{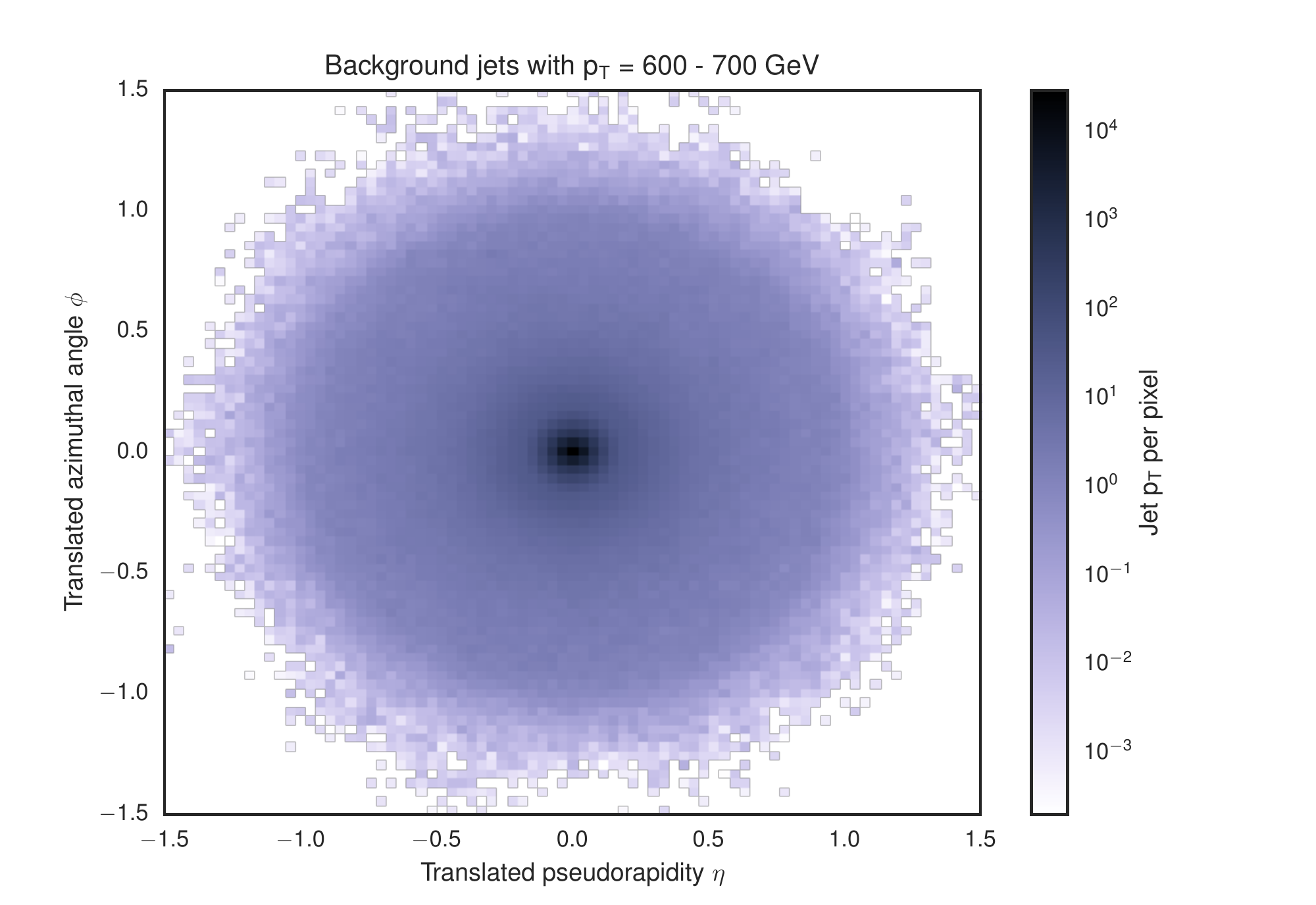}
\includegraphics[width=0.48\textwidth]{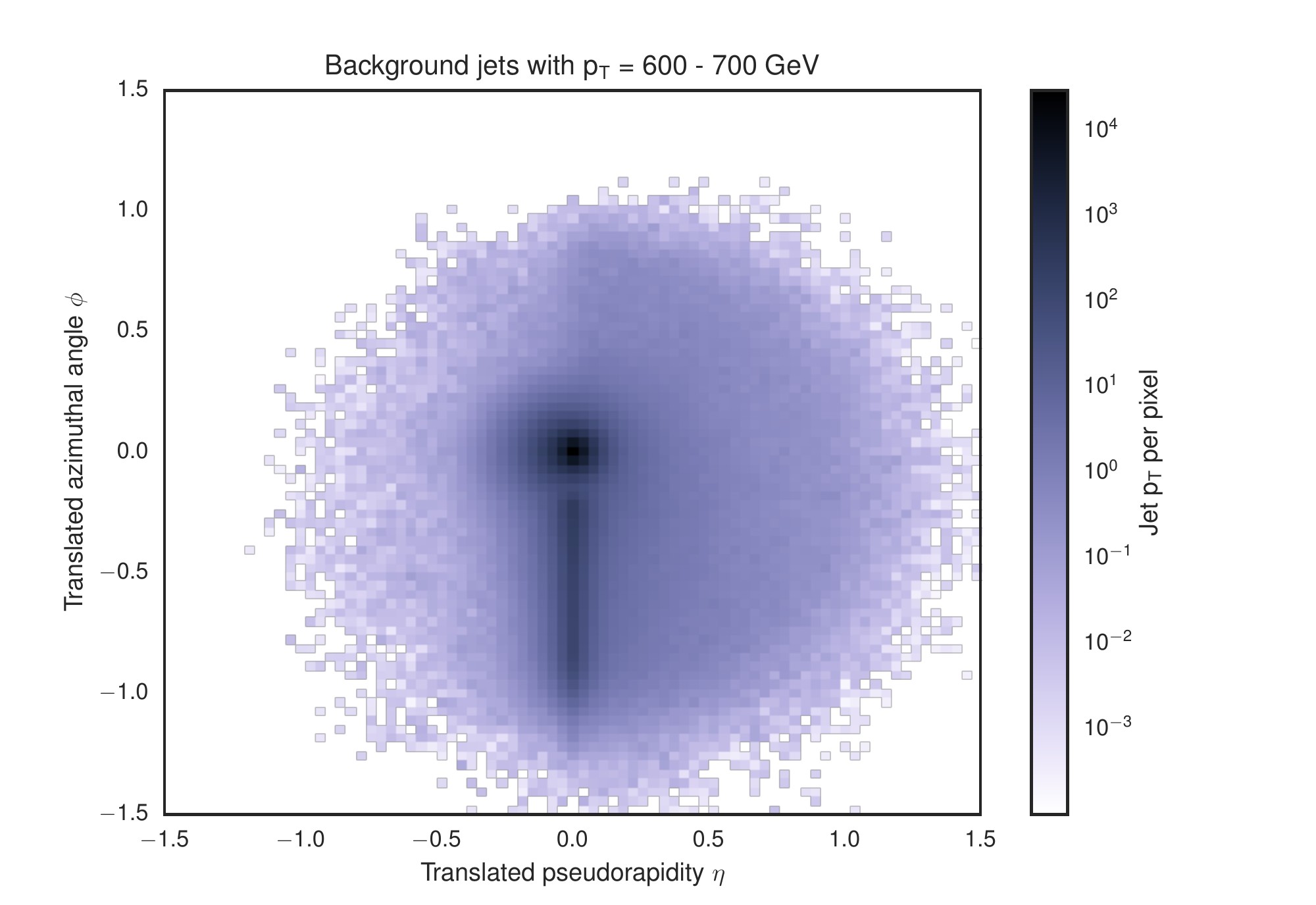}
\caption{Compound jet images after scaling and translation (left column) and after all preprocessing steps (right column) for all signal (top row) and all background (bottom row) jets. Jets with \pt\ in the range 600--700~GeV are shown.}\label{fig:jet_images}
\end{figure}

Finally, the jets are flipped in $\eta$ if the average jet \pt\ lies on the left side on the $\eta-\phi$ plane. 
Figure~\ref{fig:jet_images} shows the effect of the rotation and flip steps on $\sim$400,000 jets with \pt\ in the range between 600 and 700~GeV. These images represent the average jet image in this mass range. 
A noticeable difference between signal and background is the more densely populated 'halo' around the signal jets; this corresponds to the three-prong decay of the top quark.

The resulting jet constituents in each jet are ordered either by jet constituent \pt\ (\pt-ordering), or by subjet \pt\ and then jet constituent \pt\ (subjet ordering) and then presented to the neural network. The latter ordering is the primary choice selected in this article. As the sequences of (\pt, $\eta$, $\phi$) vary in length for each jet, the sequence was truncated at 120 jet constituents, resulting in 360 input features. This encompasses the majority of the jet constituents in each jet. Sequences were zero-padded when fewer than the maximum number of constituents are available.

%% file: sections/sec-performance-analysis.tex
\begin{figure}[htb]
\centering
\includegraphics[width=0.7\textwidth]{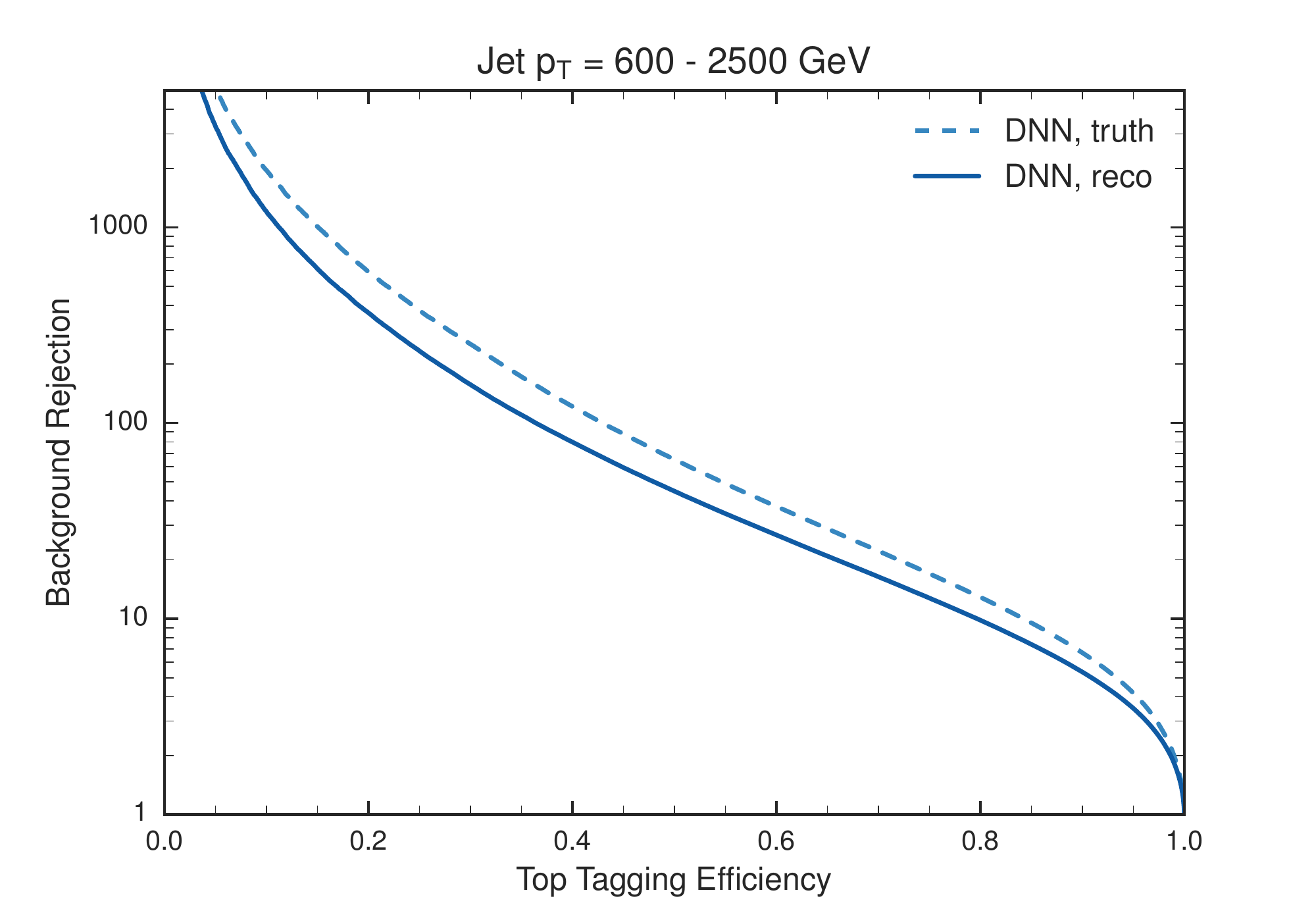}
\caption{The background rejection as a function of signal efficiency for the deep neural network. 
The performance on reconstruction level jets (solid line)
as well as on truth level (dashed line) is shown. 
For the reconstruction level jet sample, the LHC 2016 pileup scenario was used; for the truth jets, no pileup was added.}
\label{fig:performance_trimming_jss_roc} 
\end{figure}

Figure~\ref{fig:performance_trimming_jss_roc} shows Receiver Operating Characteristic (ROC) curves, for the deep neural network trained on reconstructed jets in the LHC 2016 pileup scenario as well as on truth jets without pileup. The ROC curve displays the dependency of the background rejection on the signal efficiency. Background rejection is defined as the inverse of the efficiency for accepting a background jet as signal, for a given efficiency operating point. 
This figure also shows the performance on truth jets. The deep neural network trained and tested on truth jets outperforms the one trained and tested on reconstructed jets, but seems to have a milder degradation than what has been reported elsewhere, e.g. Ref.~\cite{kyle}. 
This degradation in performance nonetheless underlines the importance of a realistic detector simulation while designing methods for large $R$-jet tagging.

While a commonly quoted measure of binary classifier performance is the Area Under the Curve (AUC), 
in a typical physics analysis, a classifier operating point or a set of points would be picked depending on the expected signal yield as well as level of background contamination. 
Table~\ref{tab:basicperf} thus shows the AUC as well as the background rejection factors obtained for 20\%, 50\%, and 80\% signal efficiency. The resampling study, described in section~\ref{sec:dataset}, found a standard deviation of $10^{-4}$ for the AUC, 4 for the background rejection at 20\% signal efficiency, 0.1 for the background rejection at 50\% signal efficiency and 0.01 for the background rejection at 80\% signal efficiency. 
Given these numbers, in table~\ref{tab:basicperf} and most subsequent tables the AUC is quoted to 3 significant figures, rejection at 20\% signal efficiency is rounded to the nearest multiple of 5, the rejection at 50\% signal efficiency is quoted to the nearest integer and the rejection at 80\% signal efficiency is rounded to 0.1.

\begin{table}[htbp]
\centering
\begin{tabularx}{0.7\textwidth}{lcYYY}

\toprule
\multirow{2}{*}{} & \multirow{2}{*}{AUC} & \multicolumn{3}{c}{Rejection at signal efficiency of} \\
 \cmidrule(lr){3-5}
 & & {20\%} & {50\%} & {80\%} \\
\hline
Reconstructed jets & 0.934 & 365 & 45 & 9.8 \\
Truth jets                & 0.946 & 595 & 65 & 12.9 \\
\bottomrule
\end{tabularx}
\caption{Area under the curve and background rejection factors for 20\%, 50\% and 80\% signal efficiency for the DNNs trained on reconstruction level and truth jets. For the reconstruction level jet sample, the LHC 2016 pileup scenario was used; for truth jets, no pileup was added.}
\label{tab:basicperf}
\end{table}

\subsection{Performance dependency on jet transverse momentum}
\begin{figure}[h!]
\centering
\includegraphics[width=0.48\textwidth]{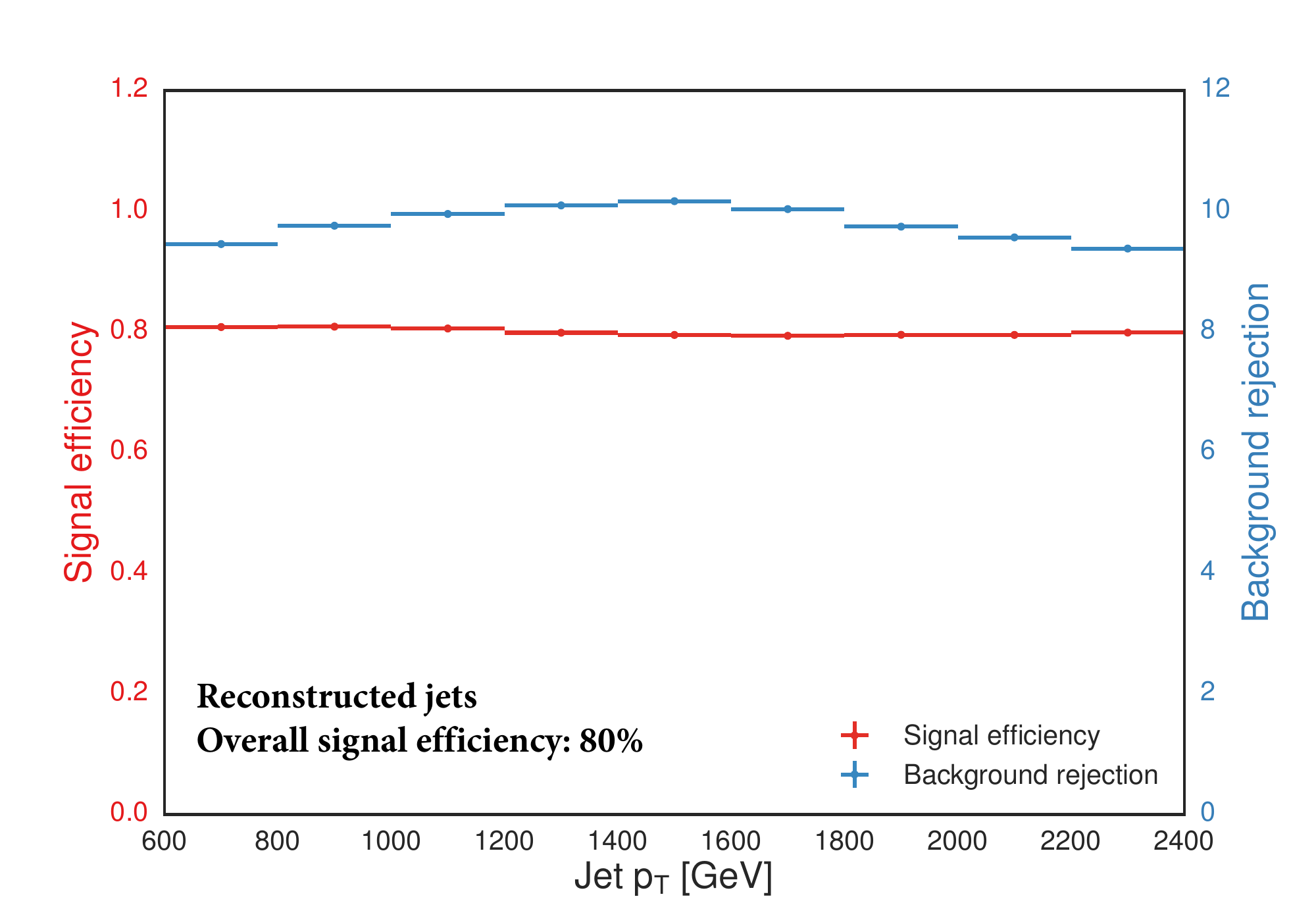}
\includegraphics[width=0.48\textwidth]{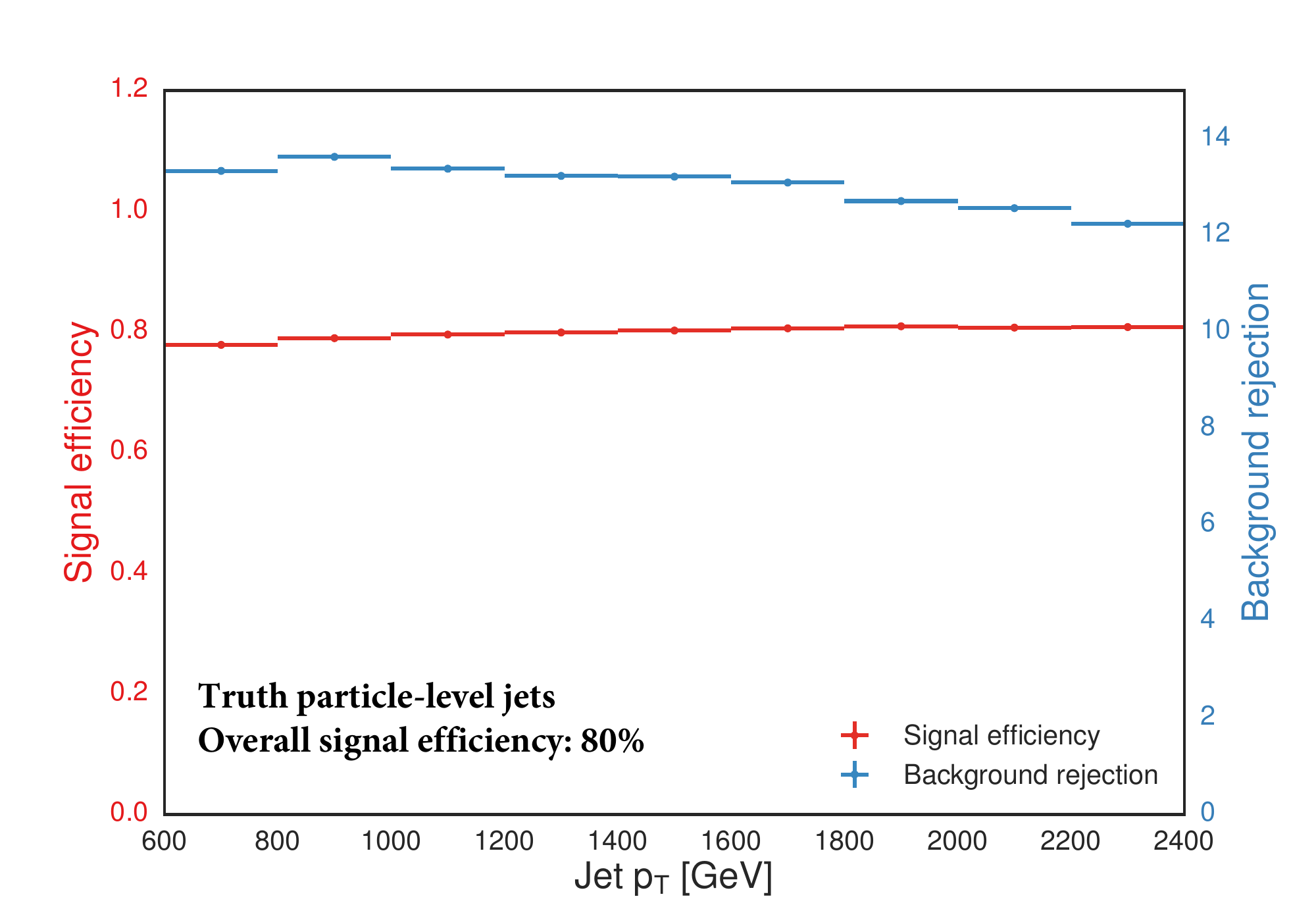}
\includegraphics[width=0.48\textwidth]{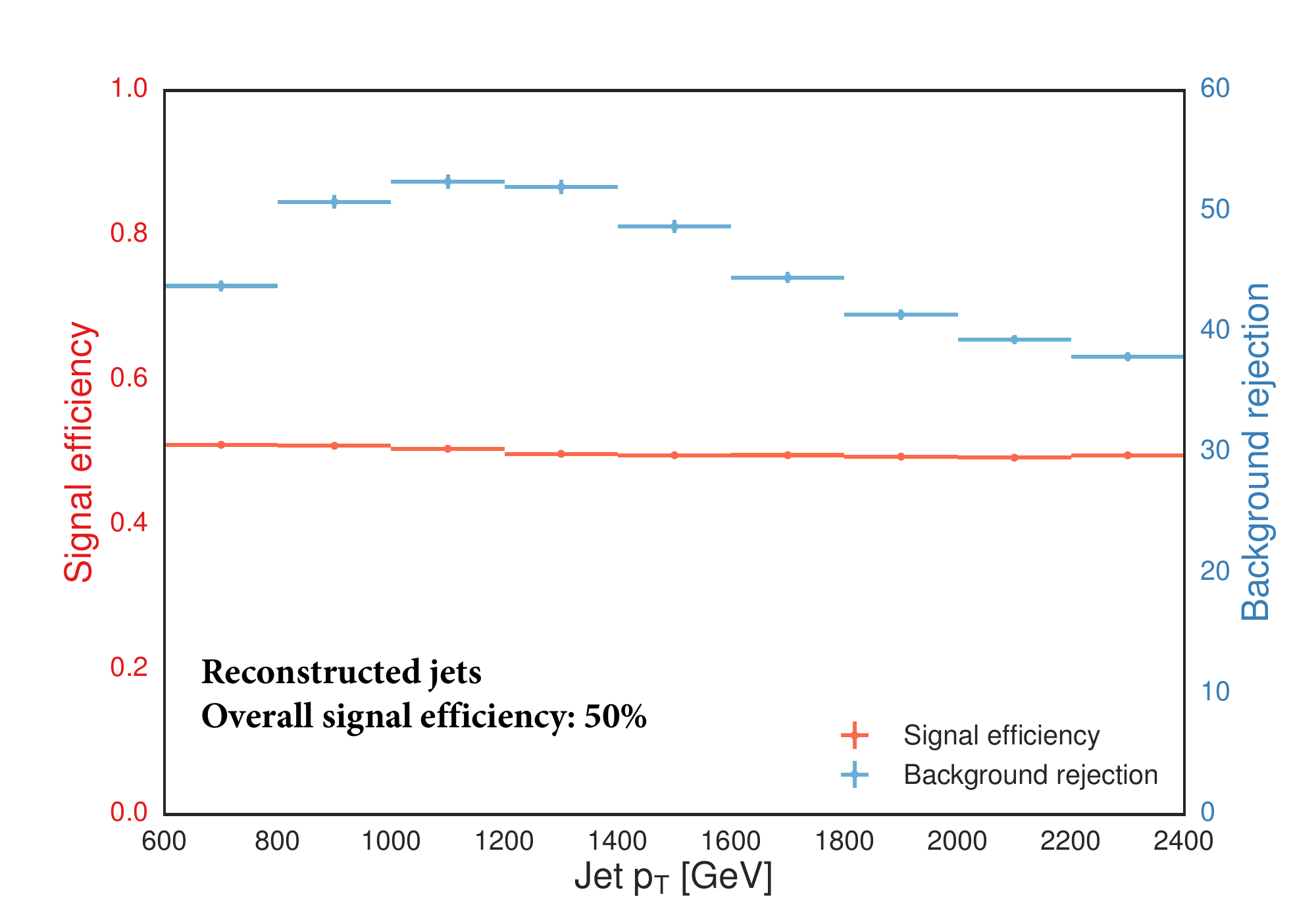}
\includegraphics[width=0.48\textwidth]{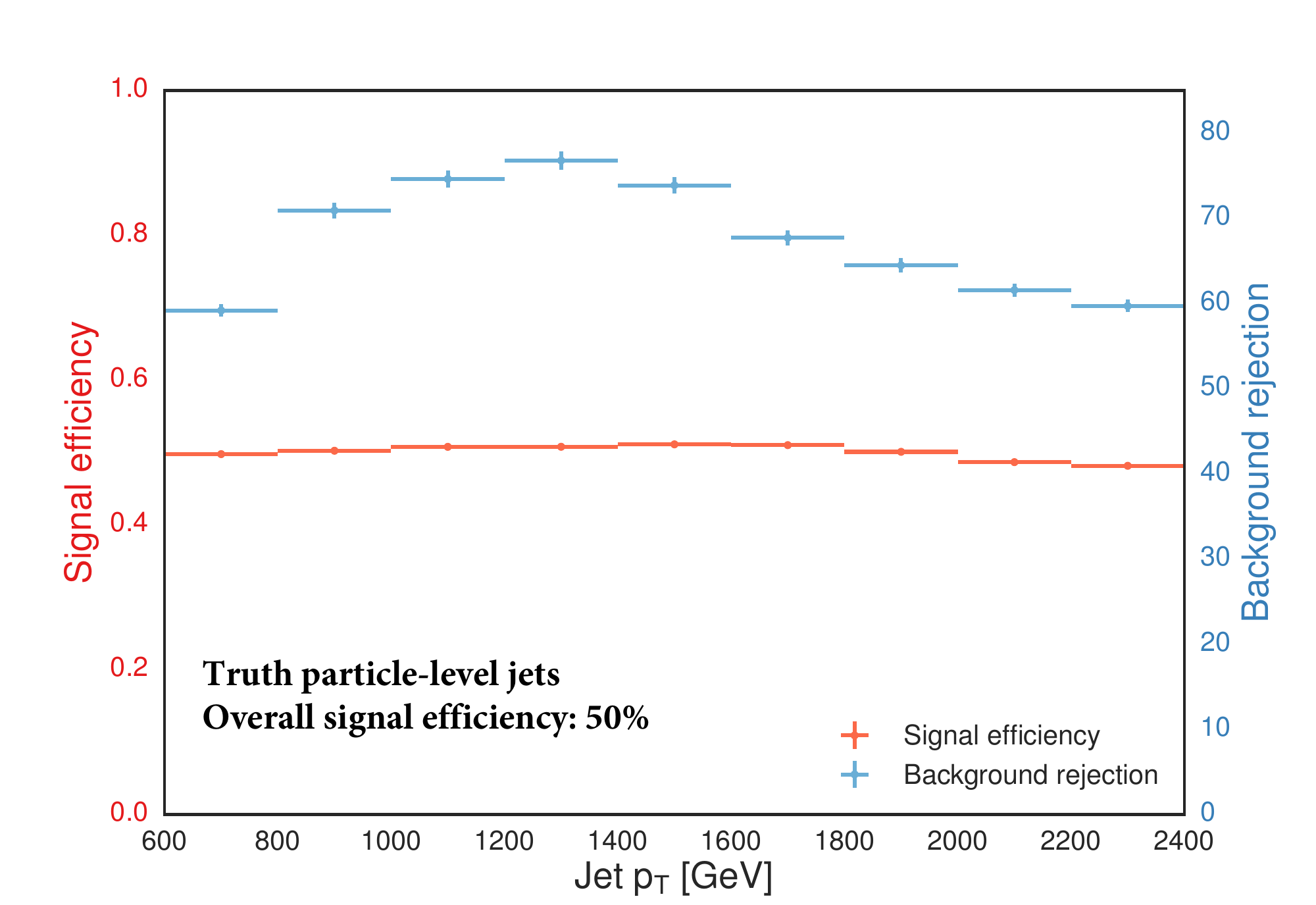}
\includegraphics[width=0.48\textwidth]{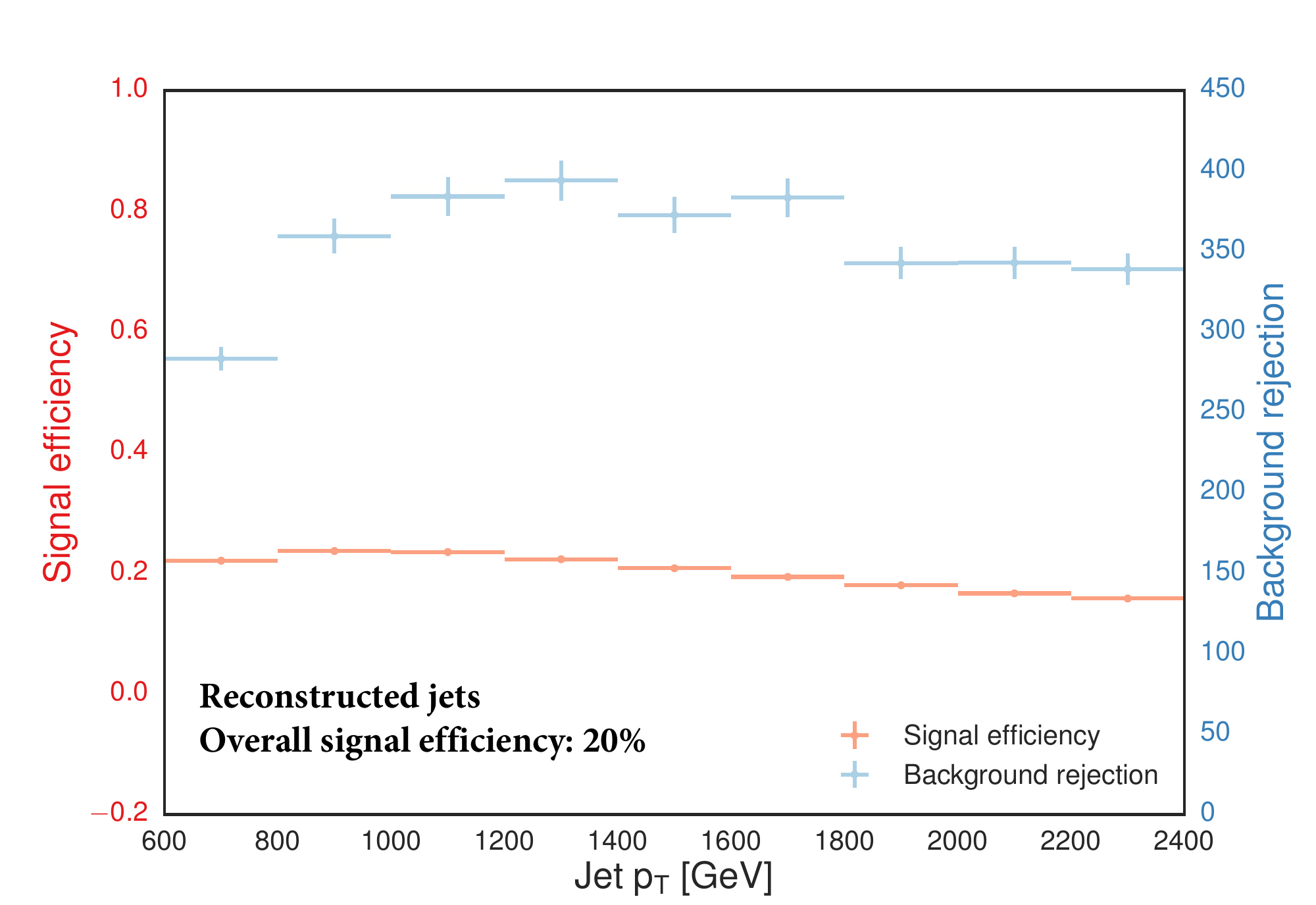}
\includegraphics[width=0.48\textwidth]{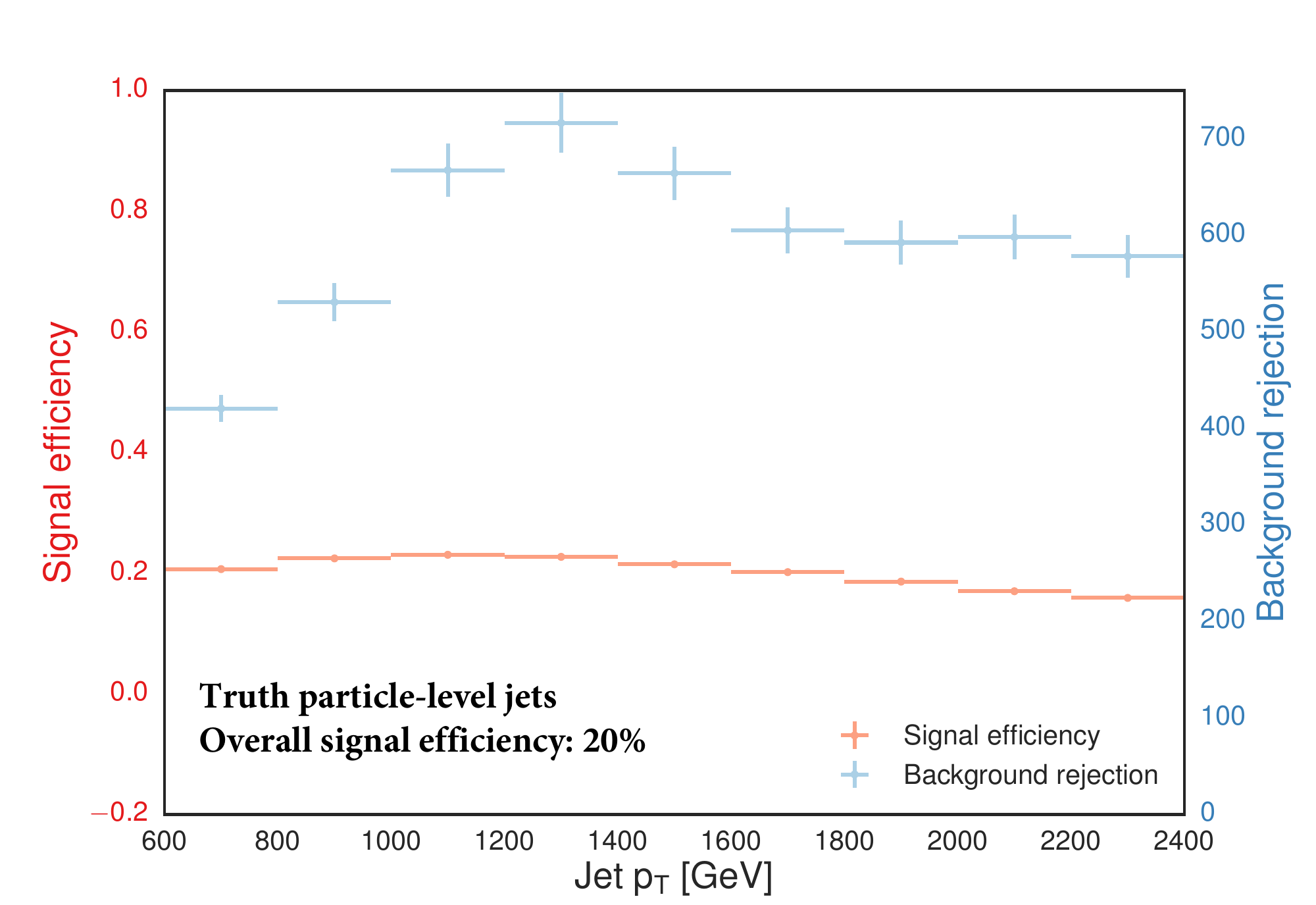}
\caption{Dependency of the top tagging efficiency (red) and background rejection (blue) on jet \pt\, for different overall signal efficiencies for reconstructed jets assuming the LHC 2016 pileup scenario (left) and truth particle jets assuming no pileup (right). Overall signal efficiency points of 80\% (top), 50\% (middle) and 20\% (bottom) are shown.}
\label{fig:sig_eff_pt_performance}
\end{figure}

The variation of the performance over the transverse momenta range studied is of particular interest to future physics analyses. 
As described previously, we attempted to mitigate the dependency on \pt\ by training using flat sampling of the \pt\ spectra for both the signal and background.
In Figure~\ref{fig:sig_eff_pt_performance} the performance at 80\%, 50\%, and 20\% overall signal efficiency is shown over the full 600--2500 GeV range, for reconstruction and truth level jets.
The deep neural network displays a remarkably flat performance in signal efficiency. 
The performance in terms of background rejection is also relatively flat, though the rejection slightly increases between 600~GeV and approximately 1~TeV and then begins to decrease. 
Two effects may explain this: first at lower momenta a small fraction of top decay products are not fully contained within the jet, and second, either the detector spatial resolution or the fixed $R$-parameter of re-clustering may be causing jets at very high \pt\ to have altered structure with respect to jets at moderate momenta. 

\clearpage
\subsection{Preprocessing Studies}\label{sec:diff_preprocessing}

\begin{figure}[h!]
\centering
\includegraphics[width=0.7\textwidth]{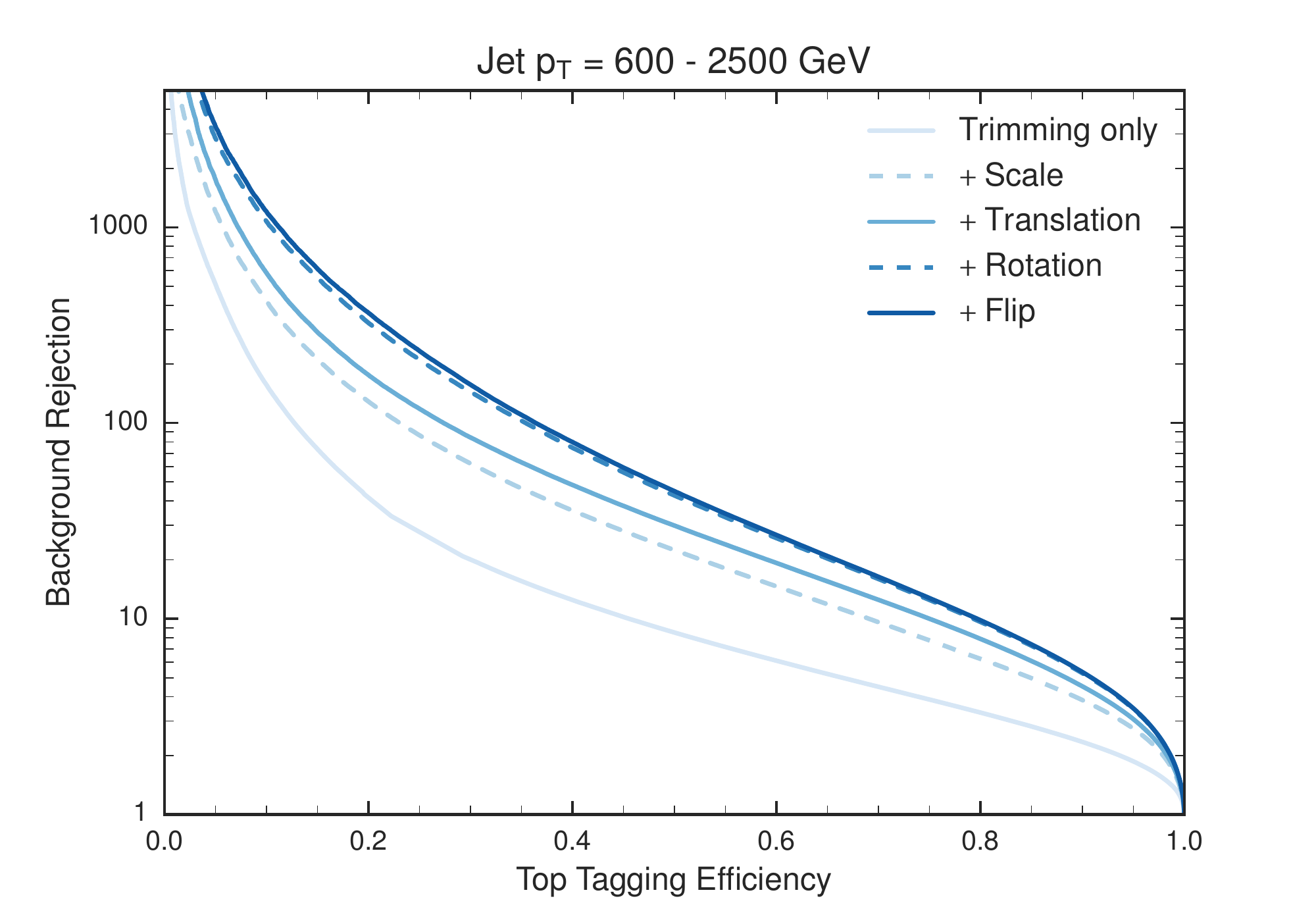}
\caption{ROC curve for DNNs trained on reconstruction level jets after each successive preprocessing step.
The LHC 2016 pileup scenario was used.
}\label{fig:atlas_roc_preprocessing_reco} 
\end{figure}

The effect of multiple different preprocessing steps were studied to optimise the tagger performance. Figure~\ref{fig:atlas_roc_preprocessing_reco} illustrates the performance gain from each sequential preprocessing step: trimming, scaling, translation, rotation and finally flipping. 
Each step has a positive impact on overall performance, with the final flipping step improving the performance only marginally. Table~\ref{tab:preprocperf} summarises the performance increase following each preprocessing stage for the AUC and rejection for the given signal efficiency operating points.

\begin{table}[htbp]
\centering
\begin{tabularx}{0.75\textwidth}{lcYYY}
\toprule
\multirow{2}{*}{Preprocessing step} & \multirow{2}{*}{\,\,\,AUC\,\,\,} & \multicolumn{3}{c}{Rejection at signal efficiency of} \\
\cmidrule(lr){3-5}
 & & \multicolumn{1}{c}{20\%} & \multicolumn{1}{c}{50\%} &\multicolumn{1}{c}{80\%} \\
\hline
Trimming only                 & 0.827 & 45  & 9  & 3.3 \\
After scaling                 & 0.904 & 130 & 22 & 6.3 \\
After translation             & 0.920 & 175 & 30 & 7.9 \\
After rotation                & 0.933 & 325 & 43 & 9.6 \\
After flip                    & 0.934 & 365 & 45 & 9.8 \\
\bottomrule
\end{tabularx}
\caption{
Area under the curve and background rejection factors for 20\%, 50\% and 80\% signal efficiency for the DNNs trained on reconstruction level jets after each successive preprocessing step. 
The LHC 2016 pileup scenario was used.}
\label{tab:preprocperf}
\end{table}

\begin{figure}[htb]
\centering
\includegraphics[width=0.7\textwidth]{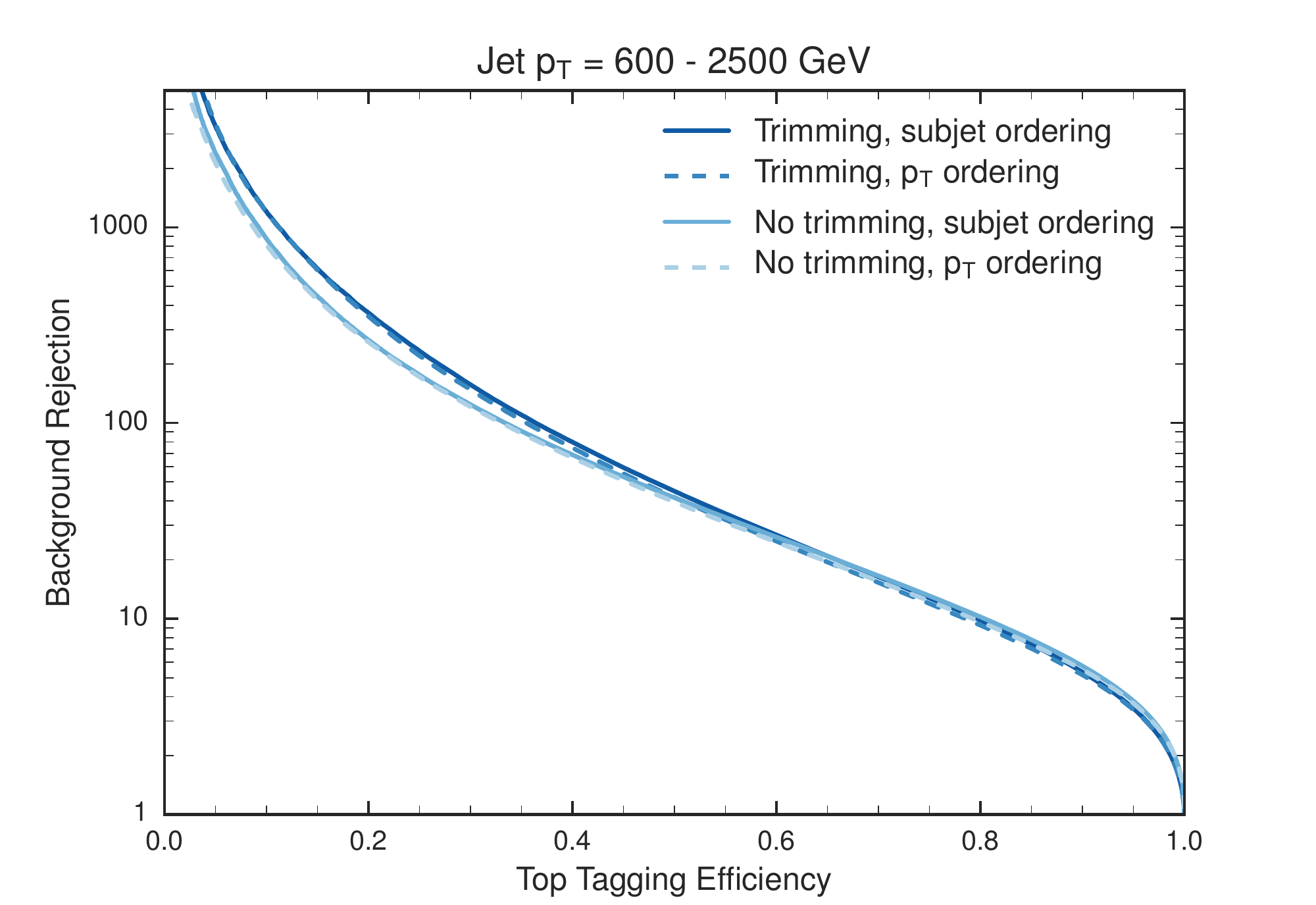}
\caption{ROC curve for DNNs trained on reconstruction level jets with different trimming and constituent ordering applied. Successive preprocessing stages (scaling, translation, rotation and flipping) are applied for all curves.  
The LHC 2016 pileup scenario was used.}
\label{fig:trimming_vs_notrim}
\end{figure}

The effect of trimming and jet constituent ordering was also investigated. Figure~\ref{fig:trimming_vs_notrim} shows the impact of the jet trimming on the ROC curve, with the same subsequent preprocessing steps applied in all cases.
Trimmed jets typically perform better at the high background rejection operating point often desired in an analysis setting. Networks trained on jets without trimming perform marginally better at the signal efficiency operating points of approximately 65\% and higher. 
The subjet or \pt\ ordering has a very small effect on the overall performance, but subjet ordering was found to have the best performance. Details of the model performance under different trimming and constituent ordering are shown in table~\ref{tab:trimsortperf}. 

\begin{table}[htbp]
\centering
\begin{tabularx}{\textwidth}{cccYYY}
\toprule
\multirow{2}{*}{Trimming} & \multirow{2}{*}{Constituent ordering} & \multirow{2}{*}{\,\,\,AUC\,\,\,} & \multicolumn{3}{c}{Rejection at signal efficiency of} \\
\cmidrule(lr){4-6}
 & & & \multicolumn{1}{c}{20\%} & \multicolumn{1}{c}{50\%} &\multicolumn{1}{c}{80\%} \\
\hline
Yes                           & \,\,\,\,\,\,\,\,\,\,\,\,\,\,\,\,\,\,subjet ordering\,\,\,\,\,\,\,\,\,\,\,\,\,\,\,\,\,\, & 0.934 & 365  & 45  & 9.8 \\
Yes                           & \pt\ ordering   & 0.931 & 350  & 42  & 9.3  \\
No                            & subjet ordering & 0.937 & 265  & 42  & 10.2 \\
No                            & \pt\ ordering   & 0.934 & 260  & 40  & 9.6 \\
\bottomrule
\end{tabularx}
\caption{Area under the curve and background rejection factors for 20\%, 50\% and 80\% signal efficiency for the DNNs trained on reconstruction level jets with or without trimming and using subjet or \pt\ ordering. Successive preprocessing stages (scaling, translation, rotation and flipping) are applied for all cases. The LHC 2016 pileup scenario was used.}
\label{tab:trimsortperf}
\end{table}

The effect of three different types of boosting was also studied. This was inspired by~\cite{melbourne} and~\cite{davis} that use the ideas of scaling and boosting the jets, respectively, to reduce the variability of jet substructure variables as a function of jet \pt\ and the ``scale'' of a jet image. In this study three approaches were tried: boosting the jet to its rest frame following~\cite{davis}, boosting the jet such that its \pt\ is identically 1000 GeV and boosting the jet so that the \pt\ of its primary subjet is always equal to the median \pt\ of the primary subjets of the top jets. None of the approaches yielded a significant improvement in performance.

\clearpage
\subsection{Pileup studies}\label{sec:pileup}
Pileup mitigation is of significant concern for the experiments at the LHC. Here the robustness of DNNs under different pileup conditions is studied. The model is trained and tested on the data in a given pileup scenario. Testing the networks on a pileup level on which it had not been trained is also studied.

\begin{figure}[h!]
\centering
\includegraphics[width=0.7\textwidth]{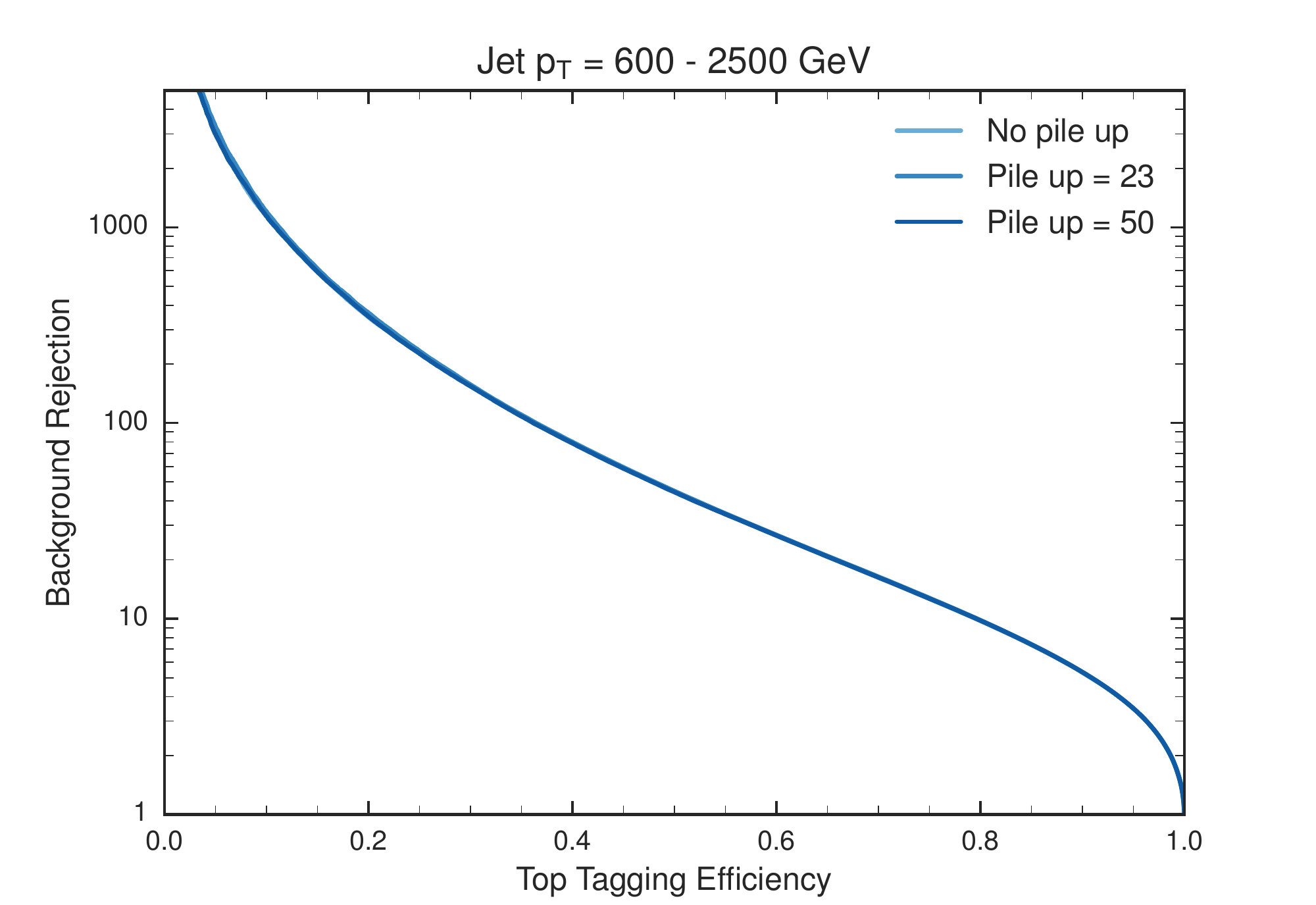}
\caption{ROC curve for DNNs trained and tested on reconstruction level jet data under three different pileup scenarios.}\label{fig:pileup_trimming_roc} 
\end{figure}

\begin{figure}[h!]
\centering
\includegraphics[width=0.7\textwidth]{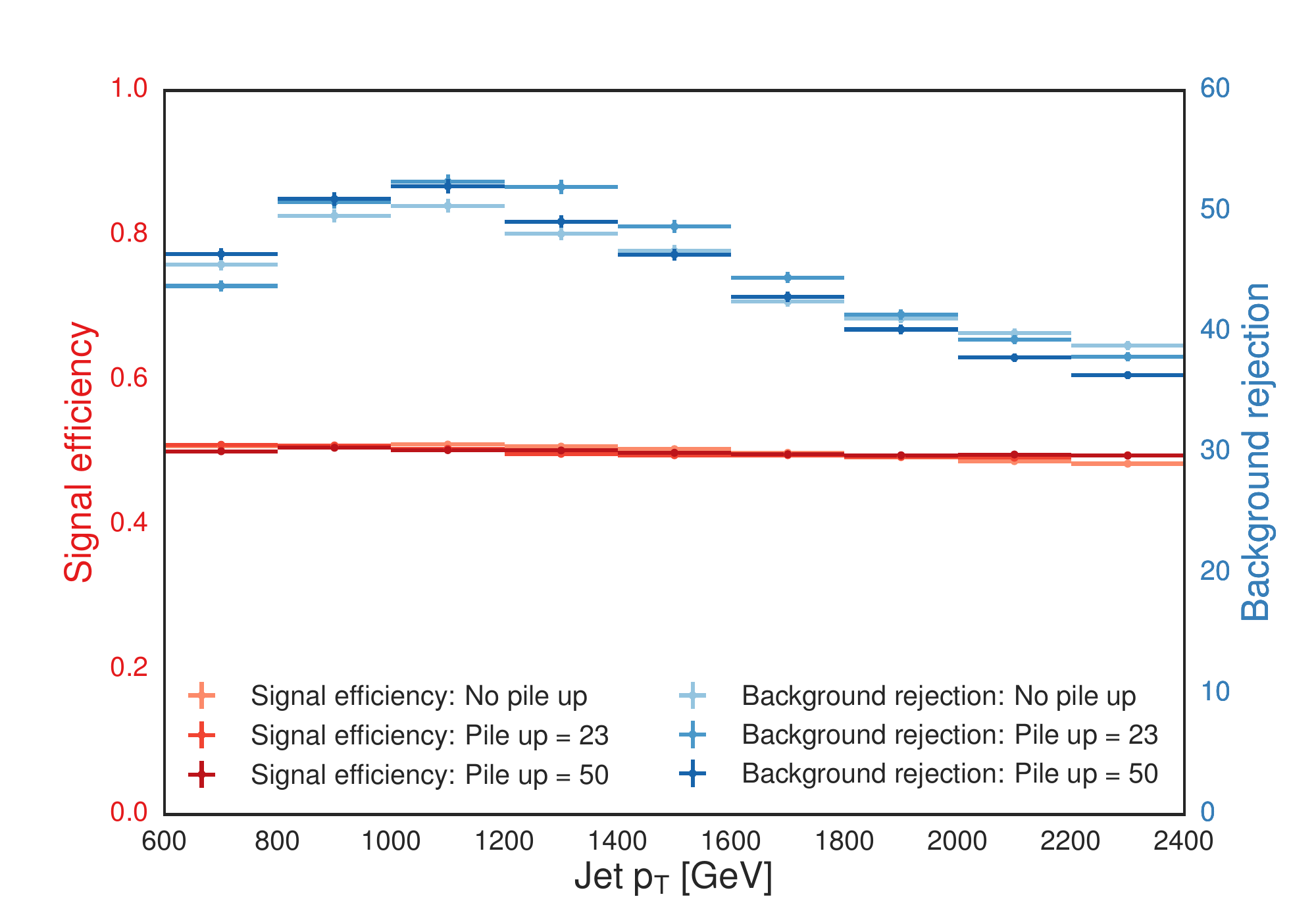}
\caption{Jet \pt\ dependency of background rejection and signal efficiency, for the overall 50\% signal efficiency working point, under three different pileup conditions.}\label{fig:pileup_trimming} 
\end{figure}

Figure~\ref{fig:pileup_trimming_roc} shows the DNN performance when training and testing on trimmed, reconstructed jets for various levels of pileup. 
The sensitivity to this level of pileup is very small, in large part due to the use of only inputs from trimmed jets.
Figure~\ref{fig:pileup_trimming} shows a comparison of the \pt\ dependency on the performance under various pileup conditions. The overall trend is that the rejection at low \pt\ is slightly better for the high pileup cases, whereas at high \pt\ it is approximately 10\% better for low pileup scenarios, though again the dependency on pileup is not significant. 

\begin{figure}[h!]
\centering
\includegraphics[width=0.48\textwidth]{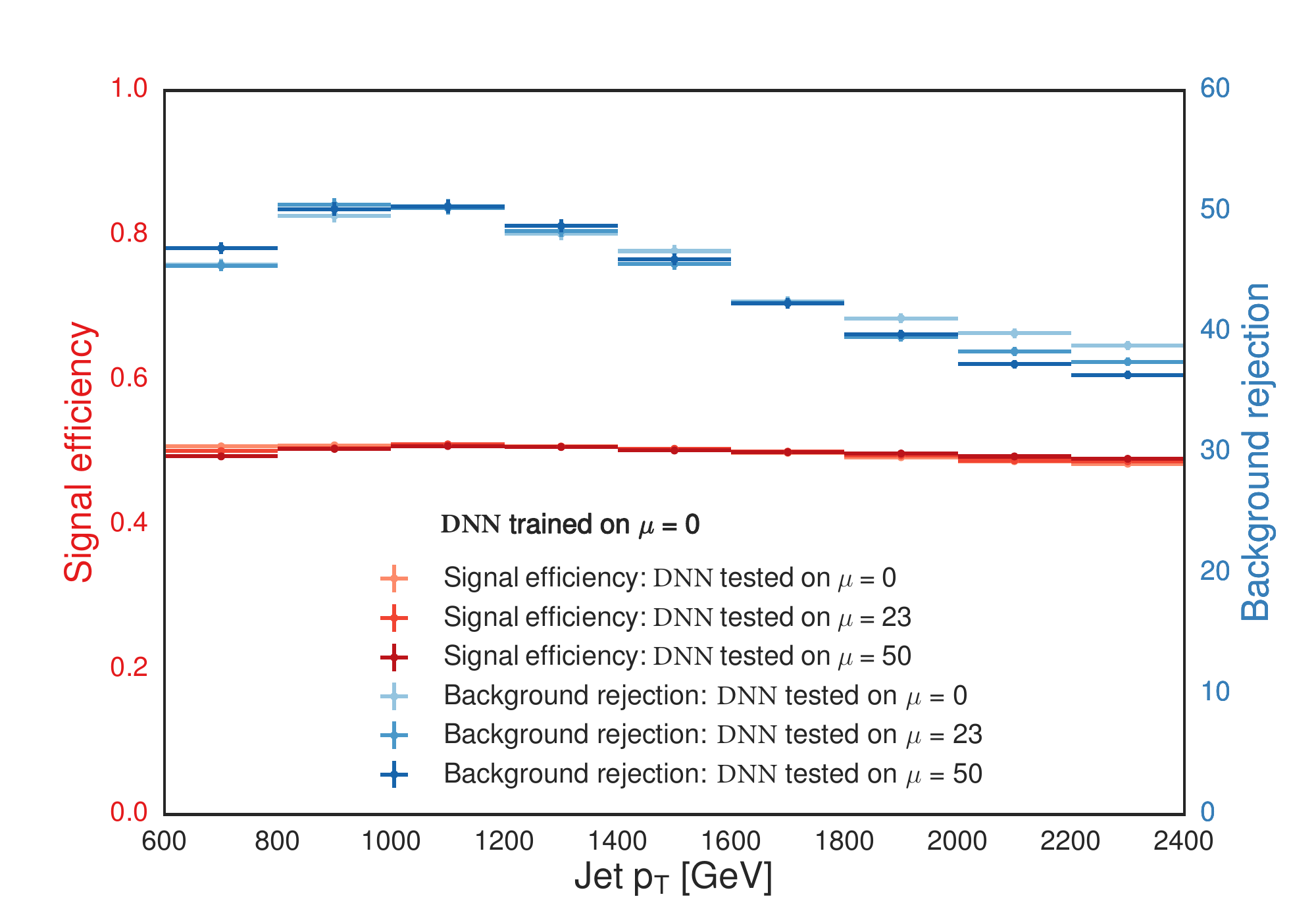}
\includegraphics[width=0.48\textwidth]{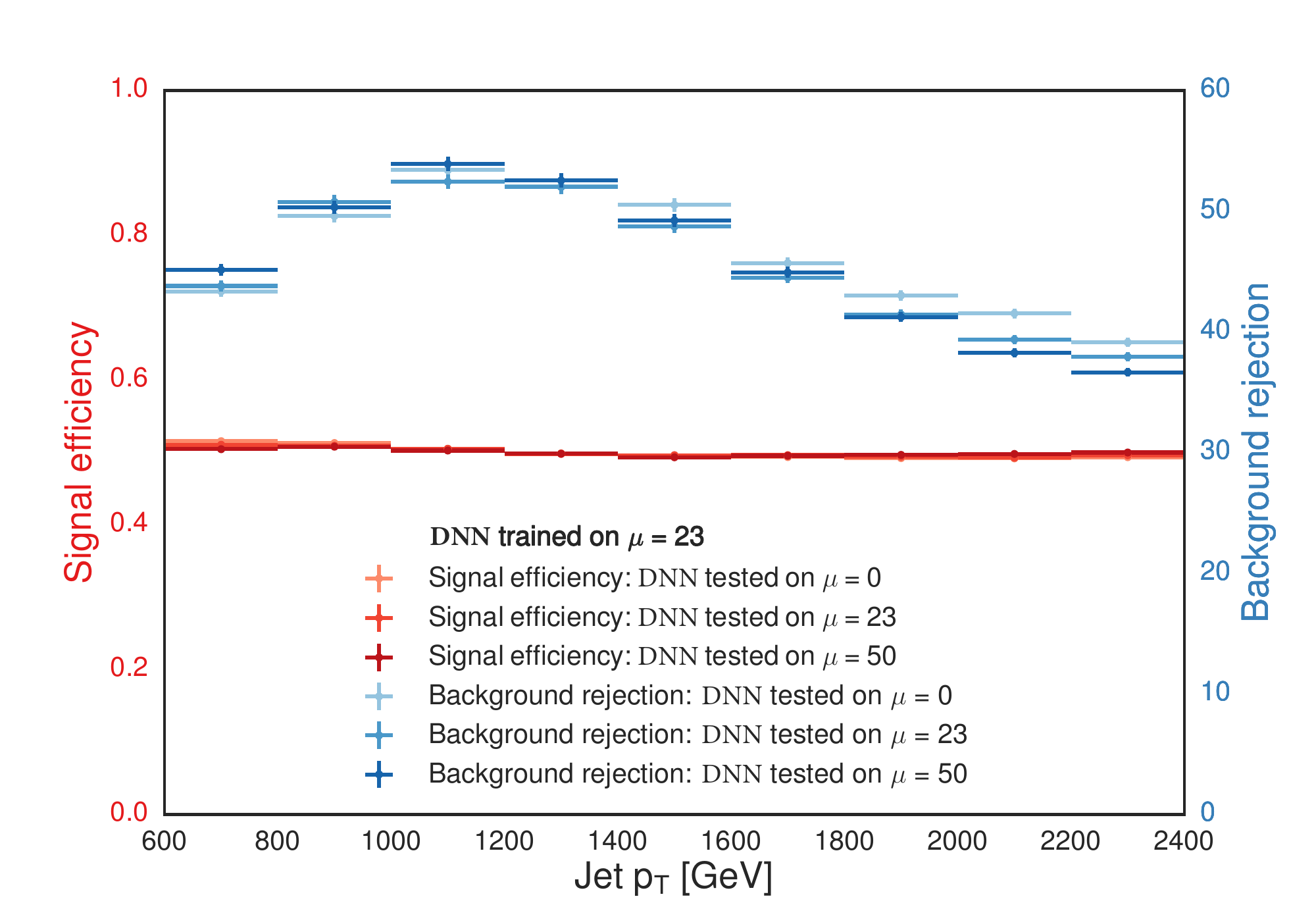}
\includegraphics[width=0.48\textwidth]{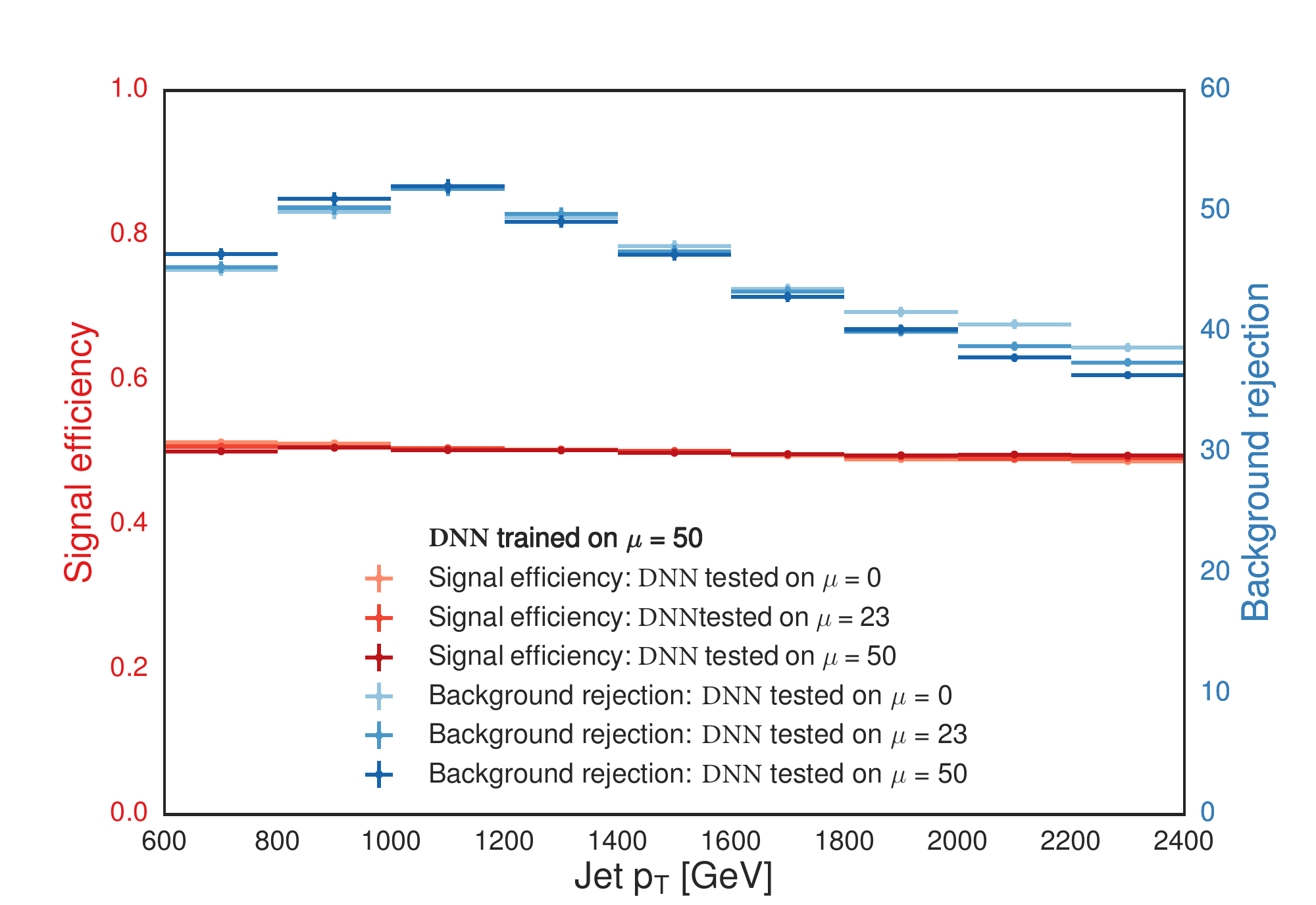}
\caption{Jet \pt\ dependency of background rejection and signal efficiency for taggers trained and tested on samples with different pileup conditions for the overall 50\% signal efficiency working point.}\label{fig:pile_up_training}
\end{figure}

Another consideration is whether the DNN would need to be retrained for different pileup scenarios. This does not appear to be the case for the pileup values expected at the LHC during Run 2.
Figure~\ref{fig:pile_up_training} shows the performance when a network is first trained on one pileup scenario, but then tested on a different scenario. The neural network again appears to be relatively robust against such variations. Indeed the overall performance is almost slightly improved for the cases with some pileup. 
A plausible hypothesis is that pileup essentially adds noise to the data. 
A common machine learning technique is to augment the data by adding noise, or using dropout~\cite{dropout} to make the DNN more robust to variations, and more able to pick out the salient features required for classification. 
Thus, deep neural networks maybe be more robust to effects like pileup which essentially mimic more noise, compared to generator or parton showering variations which can greatly affect the jet shapes~\cite{melbourne}.

\clearpage
\subsection{Performance comparison with high-level features}\label{sec:high_level_comp}
The performance of the DNN is compared to that originating from only high-level features.
In addition to comparing to the performance obtained using only the $\tau_{32}$ variable, a likelihood ratio is constructed using both the jet mass and $\tau_{32}$ observables. 
This likelihood ratio is constructed by taking two two-dimensional, uniformly binned probability density functions of $\tau_{32}$ versus jet mass, one for signal and one for the background. The discriminant is constructed as the ratio between the signal probability density function and the sum of the signal and background density functions. This method is similar to the one used in Ref.~\cite{jet_image_2} for evaluating performance of the discriminants.

\begin{figure}[h!]
\centering
\includegraphics[width=0.7\textwidth]{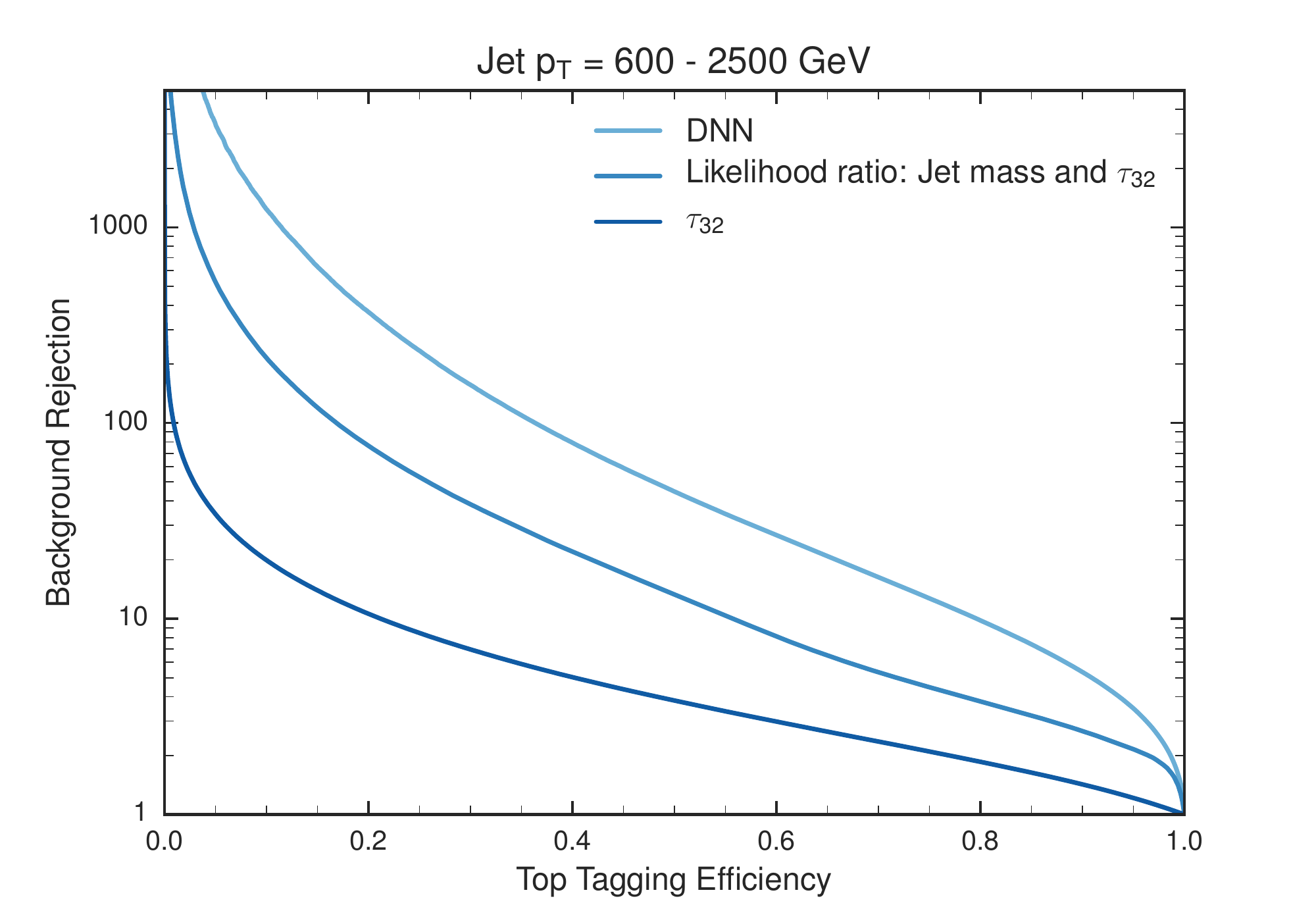}
\caption{Comparison of the ROC curves for the DNN (pale blue), a likelihood ratio discriminant built from the jet mass and $\tau_{32}$ (medium blue) and the $\tau_{32}$ variable on its own (dark blue). 
The LHC 2016 pileup scenario was used.}
\label{fig:perf_improv_high_level} 
\end{figure}

Figure~\ref{fig:perf_improv_high_level} and table~\ref{tab:perf_high_level} shows the performance comparisons, demonstrating the significant improvement obtained by the DNN.
For example for a 50\% tagging efficiency, the background rejection factor for the DNN is 3.5 times better than that of the likelihood combining $\tau_{32}$ and jet mass, and 12 times better than using the $\tau_{32}$ variable alone.

\begin{table}[htbp]
\centering
\begin{tabularx}{0.65\textwidth}{lcYYY}
\toprule
\multirow{2}{*}{Tagger} & \multirow{2}{*}{AUC} & \multicolumn{3}{c}{Rejection at signal efficiency of} \\
 \cmidrule(lr){3-5}
 & & {20\%} & {50\%} & {80\%} \\
\hline
DNN           & 0.934 & 365 & 45 & 9.8 \\
Likelihood   & 0.859 & 75 & 13 & 3.8 \\
$\tau_{32}$ & 0.678 & 11 & 3.8 & 1.9 \\
\bottomrule
\end{tabularx}
\caption{Performance metrics for the DNN, the likelihood ratio discriminant composed of $\tau_{32}$ and jet mass, and the $\tau_{32}$ alone. 
The LHC 2016 pileup scenario was used.}
\label{tab:perf_high_level}
\end{table}

\begin{figure}[h!]
\centering
\includegraphics[width=0.55\textwidth]{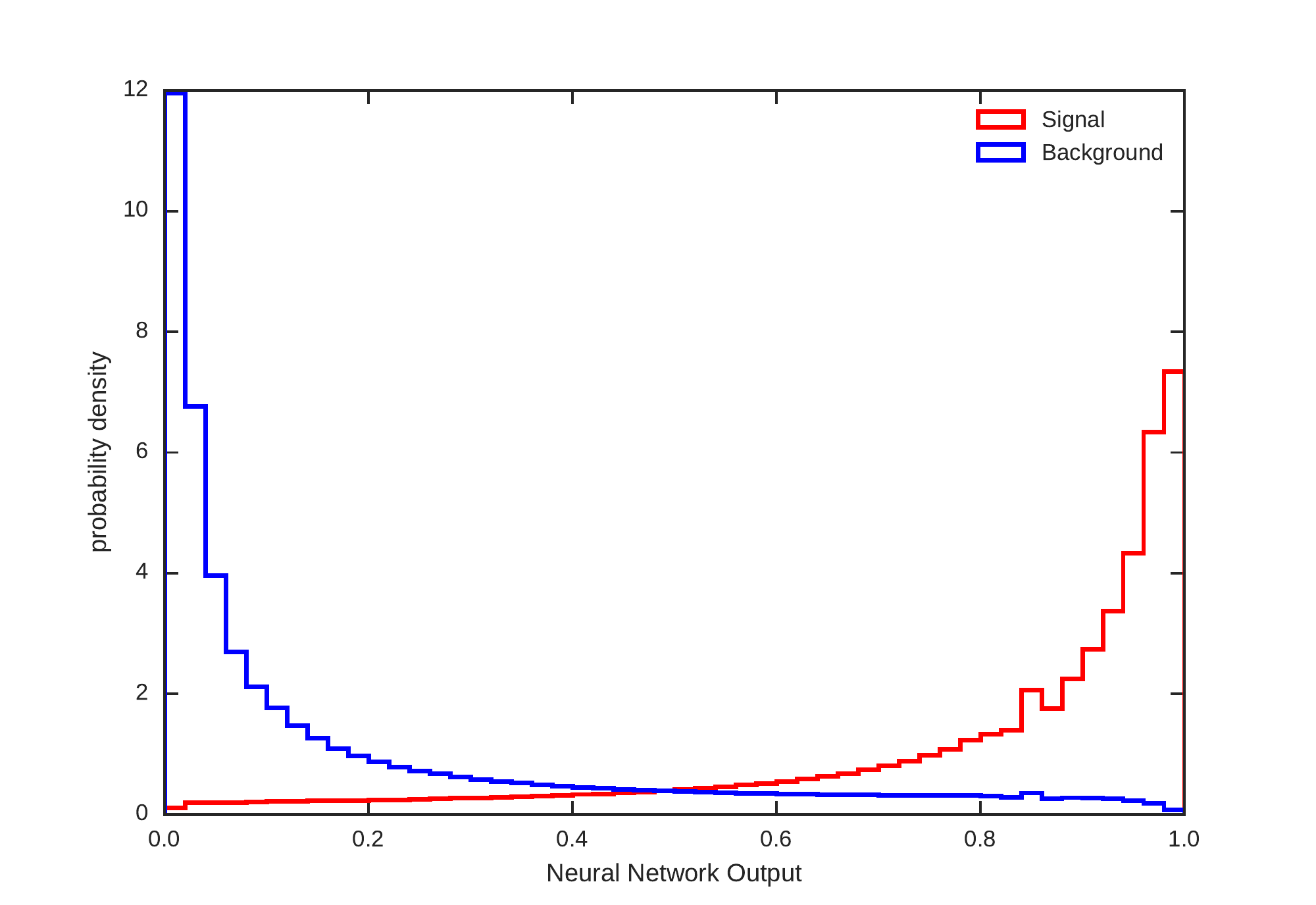}
 \includegraphics[width=0.48\textwidth]{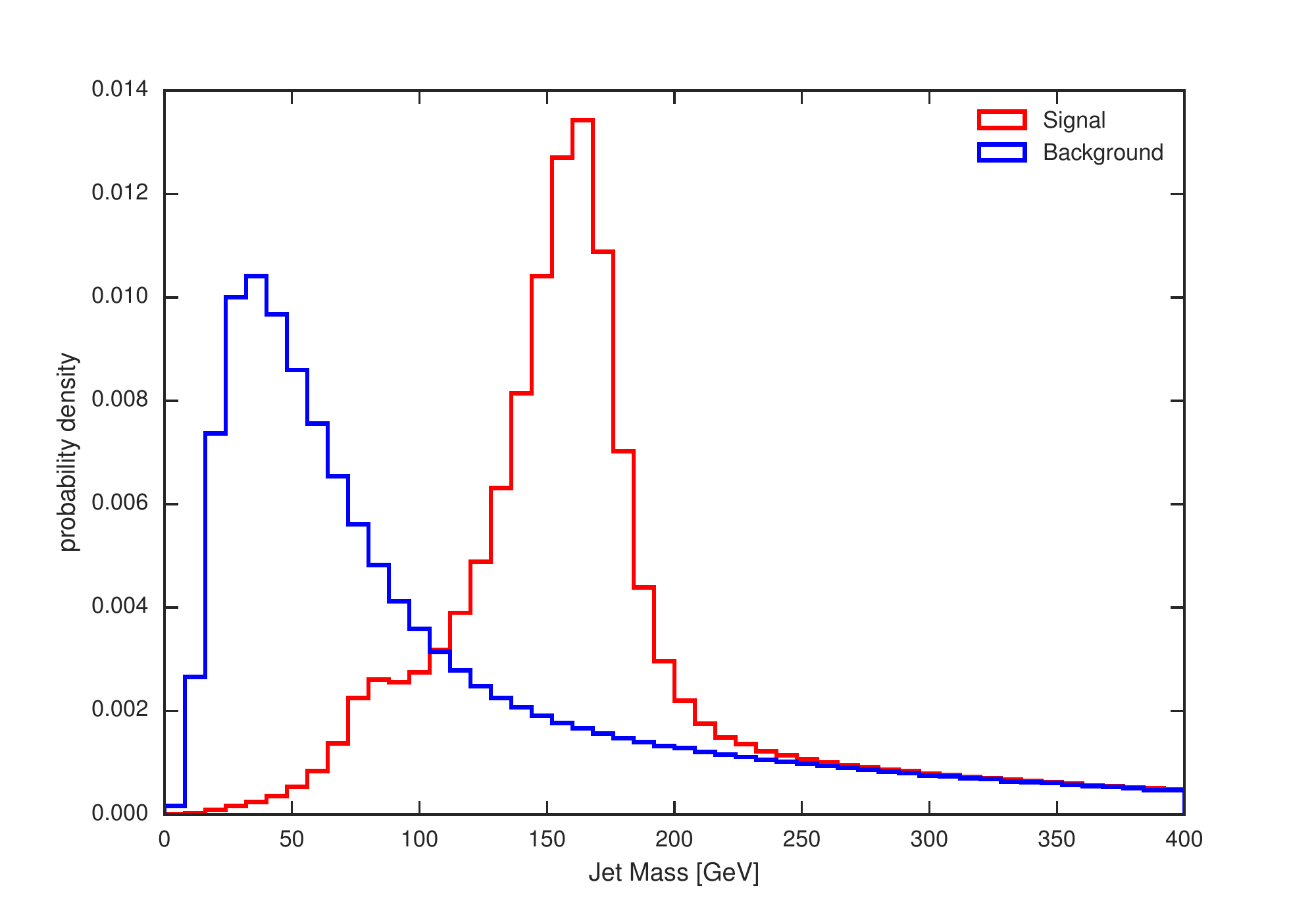}
\includegraphics[width=0.48\textwidth]{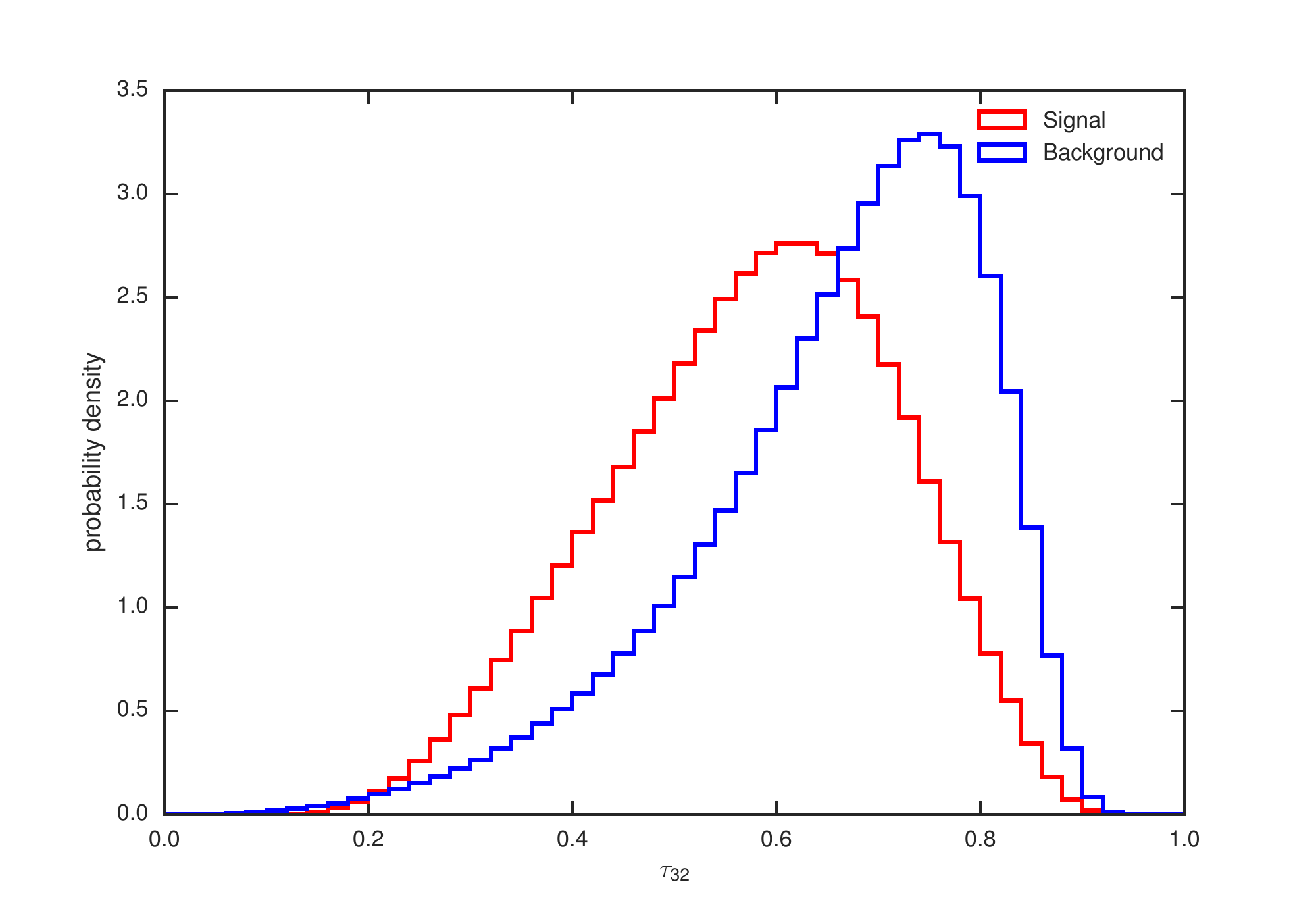}

\caption{Distribution of the DNN (top), jet mass (bottom left) and $\tau_{32}$ (bottom right) for the signal (red) and background (blue) samples. The LHC 2016 pileup scenario dataset is used.}\label{fig:high_level_1d}
\end{figure}

Figure~\ref{fig:high_level_1d} shows the distributions of the DNN output, jet mass and $\tau_{32}$ for the signal and the background. This clearly shows, for the DNN, the signal peaking at 1 as expected, while the background peaks at zero. 
The separation improvement of the DNN over the jet mass and $\tau_{32}$ high-level variables is also apparent.

\clearpage
\subsection{Correlations with high-level features}\label{sec:high_level_corr}

\begin{figure}[h!]
\centering                                                                                                          
\includegraphics[width=0.35\textwidth]{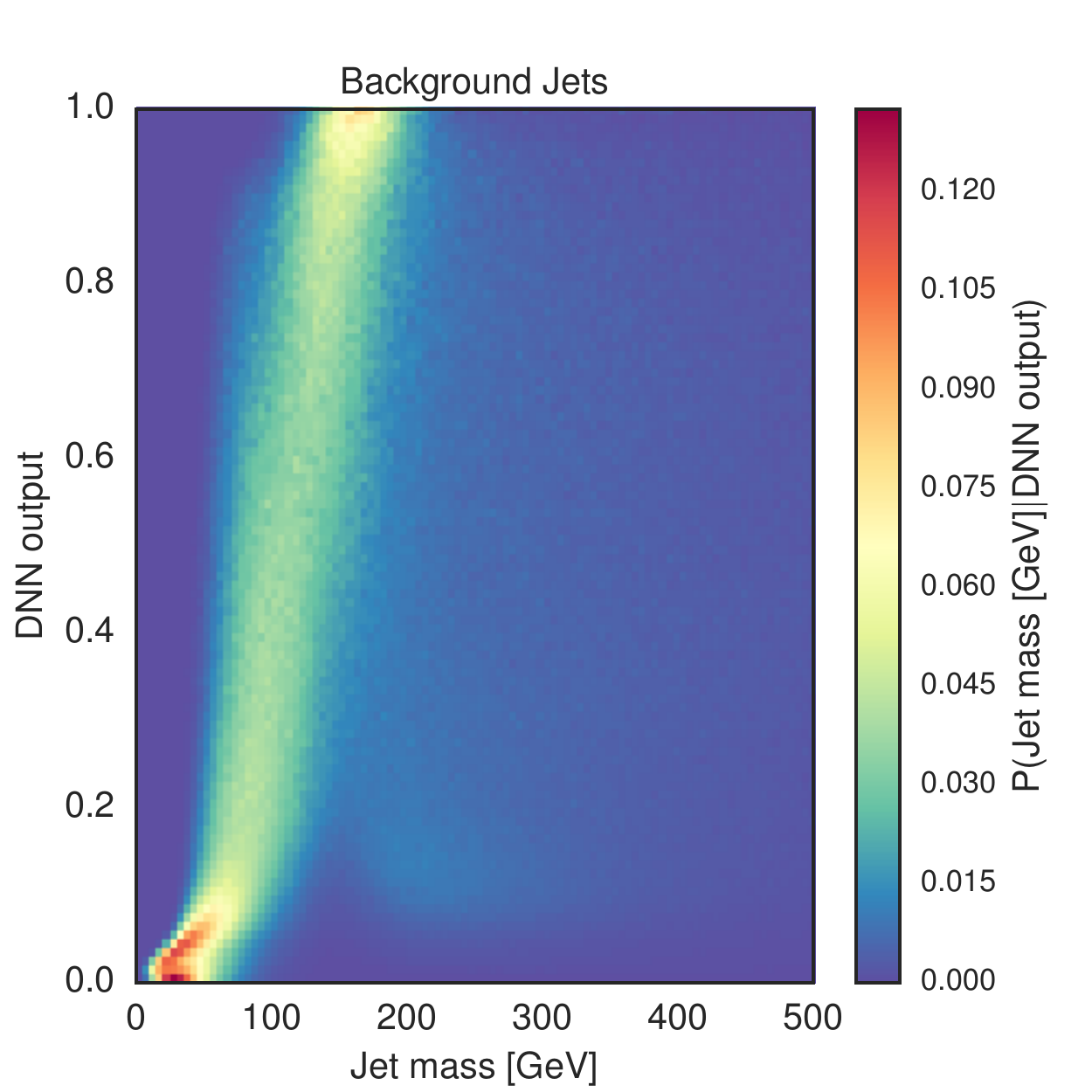}                                                                                              
\includegraphics[width=0.35\textwidth]{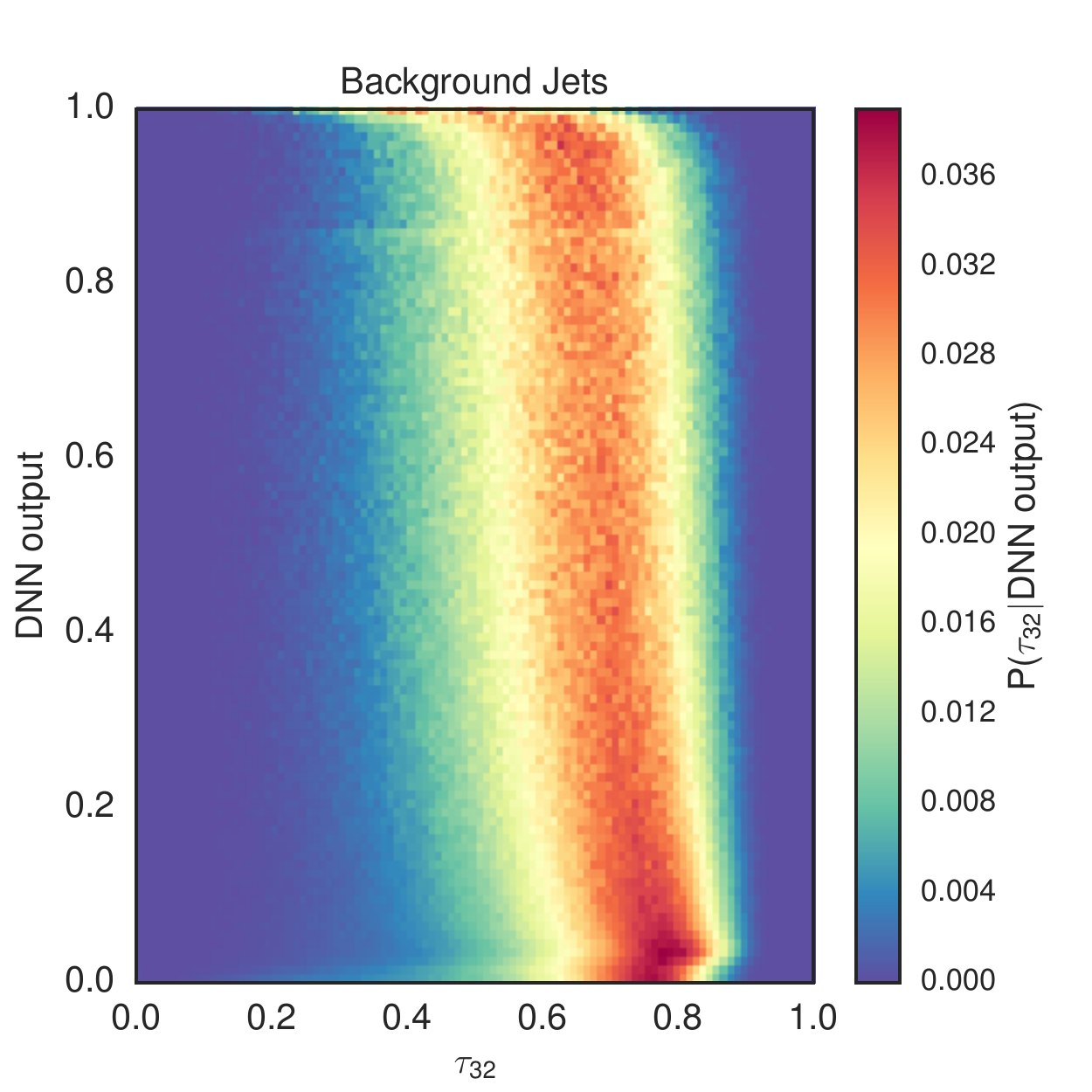}
\includegraphics[width=0.35\textwidth]{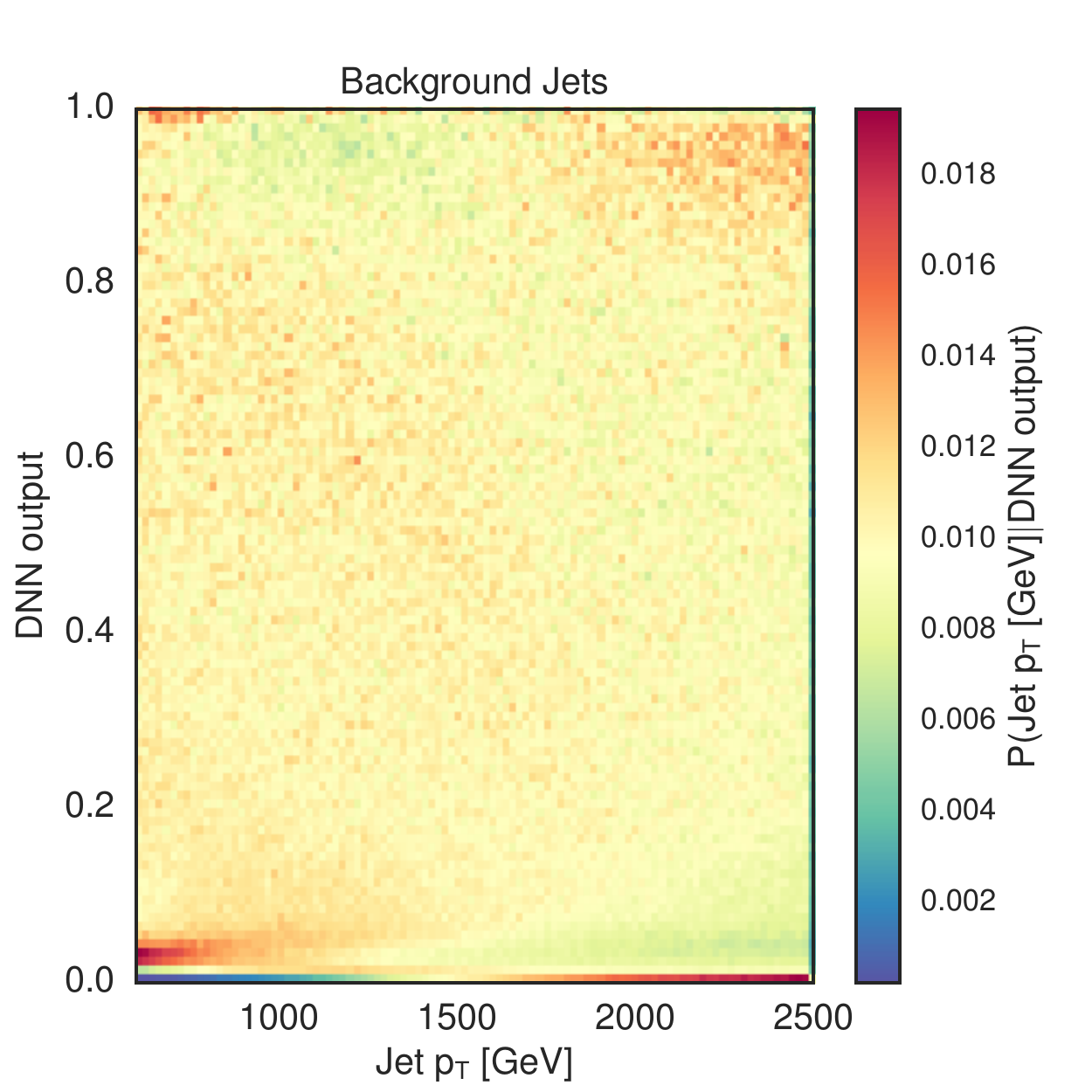}                                                  
\caption{Conditional distribution of high-level jet features on the DNN output. The top left plot shows the jet mass,
top right the $\tau_{32}$ 
and bottom plot shows the jet \pt. 
The rows have been normalised to unity. 
The LHC 2016 pileup scenario dataset is used.}
\label{fig:high_level_struc} 
\end{figure}

A qualitative evaluation is made to see if the network has learned to recognise some of the high-level features known to be important in the classification. 
Figure~\ref{fig:high_level_struc} shows the conditional distribution of the jet mass, $\tau_{32}$ and jet \pt{} as a function of the network output.   The distribution is shown for jets from the background sample
The distributions are presented as two dimensional histograms with the feature of interest on the abscissa and network response on the ordinate, where the rows of the histograms have been normalised to unity.

The conditional dependency of the jet mass on the network output shows that for the network to classify a background jet as signal-like (output close to 1.0) the jet mass has to be within approximately 30 GeV of the top quark mass. One can see a correlation between the DNN output and the jet mass, showing that network seems to have 'learned' the jet mass.
A similar yet weaker behaviour is seen in the correlation between the DNN and the $\tau_{32}$ observable. 
These are expected behaviours, given the  separation power between signal and background jets that the jet mass and $\tau_{32}$ provide.
On the other hand, as desired and designed, there is no similar relationship between the network output and the jet \pt, confirming that the network is not learning to discriminate between signal and background based on the \pt\ of the jets.

\clearpage
\subsection{Architecture studies}\label{sec:arch_studies}
Network architecture studies were performed within the fully connected network model. Several ``rectangular'' networks were tried where the number of nodes in a hidden layer was the same for each hidden layer. The depths of such networks varied from 4 to 6 hidden layers with 400 to 1000 nodes in each hidden layer. A larger ``tapered'' architecture was also tried with one more hidden layer and larger number of nodes per layer (600, 500, 300, 150, 50) with respect to the default network. None of these architectures improved the performance.

Dropout regularisation was also attempted on the default network as well as on the larger tapered network. Dropout was applied only on the input layer or  only on the hidden layers (with equal dropout probability in each hidden layer) or simultaneously on input and hidden layers with the same dropout probability. Dropout probabilities were varied from 2\% to 80\% producing no improvement on the performance.

%% file: sections/sec-conclusions.tex
In this article a method for boosted top quark jet tagging was developed. The method is based on processing a sequence of four vectors of the jet constituents and achieves a background rejection of 45 at the 50\% efficiency operating point for reconstruction level jets in the \pt\ range between 600 and 2500 GeV. The rejection achieved for truth particle jets is 65 at the 50\% efficiency operating point. Input ordering and data preprocessing methods preserving jet properties were developed and their importance in achieving high background jet rejection demonstrated. Pileup at the levels expected during Run 2 of the LHC was found not to substantially influence the performance of the classifier. Several methods of jet boosting were found not to further improve the DNN's performance. A survey of fully connected network architectures and dropout regularisation settings was conducted without showing any performance increase. 

This method can be extended in the future to incorporate Recurrent Neural Networks with Long Short-Term Memory~\cite{lstm} that are well-suited for sequence processing. Future directions for this research include the investigation of classifier sensitivity to systematic effects such as changes to the Monte Carlo generators and parton showers and applying recently developed systematic mitigation methods that incorporate adversarial training~\cite{pivot}. 
The ultimate goal is the development of a tagger usable in an experimental setting that would have increased performance relative to existing top taggers. This would result in a higher sensitivity to physics beyond the Standard Model and improved measurements of Standard Model processes with highly boosted top quarks.